\documentclass{aa} 
\usepackage{morefloats}
\usepackage{graphics}
\begin{document}
\thesaurus{11(11.01.2; 11.09.1 \object{NGC 6951}; 11.11.11; 11.14.1; 11.19.2; 11.19.6)}

\title{Circumnuclear structure and kinematics in the active galaxy \object{NGC 6951}
\thanks{Based on observations made with the 4.2m William Herschel Telescope operated by 
the Isaac Newton Group, and the 2.6m Nordic Optical Telescope operated jointly by Denmark, 
Finland, Iceland, Norway, and Sweden, on the island of La Palma in the Spanish Observatorio 
del Roque de los Muchachos of the Instituto de Astrof\'\i sica de Canarias.}
\thanks{Also based on observations made with the NASA/ESA Hubble Space Telescope, 
obtained from the data archive at the ESA Space Telescope European Coordinating Facility.}}

\author{
  Enrique P\'erez  \inst{1}
\thanks{Visiting astronomer, German-Spanish Astronomical Centre, Calar Alto, operated by 
the Max-Planck-Institute for Astronomy, Heidelberg, jointly with the Spanish National 
Comission for Astronomy.}
\and
  Isabel M\'arquez \inst{1}  
\and
  Ignacio Marrero  \inst{1} 
\and
  Florence Durret  \inst{2}
\and
  Rosa M. Gonz\'alez Delgado \inst{1}
\and
  Josefa Masegosa  \inst{1}
\and 
  Jos\'e Maza      \inst{3}
\and 
  Mariano Moles    \inst{4}
                   \thanks{On sabbatical at Queen Mary and Westfield College, Astronomy Unit, London, UK}
}
\offprints{Enrique P\'erez} 
\mail{eperez@iaa.es}
\institute{
   Instituto de Astrof\'\i sica de Andaluc\'\i a (CSIC), Apartado 3004, 
   18080 Granada, Spain
\and
   Institut d'Astrophysique de Paris, CNRS, Universit\'e Pierre et 
Marie Curie, 98bis Bd Arago, 75014 Paris, France 
\and 
   Departamento de Astronom\'\i a, Universidad de Chile, Casilla 36-D, Santiago, Chile
\and
   Instituto de Matem\'atica y F\'\i sica Fundamental (CSIC), Madrid, Spain,
and Observatorio Astron\'omico Nacional, Madrid, Spain
}
\date{Received  / Accepted }

\titlerunning{Structure and kinematics of \object{NGC 6951}}
\authorrunning{Enrique P\'erez et al.}
\maketitle
 
\begin{abstract}

A study is presented of the central structure and kinematics of the galaxy NGC~6951, 
by means of broad band {\it B{\arcmin}IJK} images and high resolution high dispersion 
longslit spectroscopy, together with archival HST WFPC2 {\it V} and NICMOS2 {\it J} 
and {\it H} images. 
We find that there is little ongoing star formation inside the bar 
dominated region of the galaxy, except for the circumnuclear ring at 5 arcsec radius. 
There is some evidence that this star formation occurs in two modes, in bursts and 
continuously, along the ring and inwards, towards the nucleus. The equivalent width 
of the \ion{Ca}{ii} triplet absorption lines show that, in the metal rich central 
region, the continuum is dominated by a population of red supergiants, while 
red giants dominate outside.  
The gaseous kinematics along three slit position angles, and the comparison with
the stellar kinematics, suggest the existence of a hierarchy of disks within disks, 
whose dynamics are decoupled at the two inner Linblad resonances (ILR), that we find 
to be located at 180 pc and at 1100 pc. This is supported by the structure seen in the 
high resolution HST images. The nucleus is spatially resolved in the emission line 
ratio [\ion{N}{ii}]/H$\alpha$, and in the FWHM of the emission lines, within a radius 
of 1.5{\arcsec}, just inside the innermost ILR. Outside the iILR, the stellar CaT 
velocity profile is resolved into two different components, associated with the bar 
and the disk. Several results indicate that this is a dynamically old system: 
the little ongoing star formation inside the bar dominated 
part of the galaxy, the very large relative amount of molecular to total mass within 
the inner 6 arcsec radius, $\sim25$\%, and the geometry of the circumnuclear ring that 
leads the stellar bar at a position angle greater than 90{\degr}. It is thus possible that 
a nuclear bar has existed in NGC~6951 that drove the gas towards the nucleus, as in the 
bars within bars scenario, but that this bar has already dissolved by the gas accumulated 
within the circumnuclear region. We discuss the possibility that the kinematical 
component inside the iILR could be due to a nuclear outflow produced by the combined 
effects of SN and SN remnants, or to a nuclear disk, as in the disk within disk scenario 
that we propose for the fueling of the AGN in NGC~6951. 

\keywords{galaxies: active -- galaxies: individual: \object{NGC 6951} -- galaxies: 
kinematics and dynamics -- galaxies: nuclei -- galaxies: spiral -- galaxies: structure}

\end{abstract}

 
\section{Introduction}

The centers of spiral galaxies are frequent sites of activity, marked by the
presence of intense star formation (SF) and/or an active galactic nucleus (AGN). 
This activity needs to be fueled with a supply of gas, whose reservoir can be provided
by the disk of spiral galaxies. An efficient way for this gas to lose its angular momentum
is provided by the existence of non-axisymmetric components of the galactic potential,
together with a strong central gravitational potential. The latter is
necessary to explain that AGNs, irrespective of the presence of morphological
perturbations, appear preferentially in early type spirals, with the frequency
peak at Sb. On the other hand, galaxies dominated by disk star formation occur
preferentially in later types with a distribution peaking at Scd (Moles et al.
1995; Ho et al. 1997a  and references therein). 

The existence of gas to fuel the circumnuclear activity is necessary
but not sufficient (e.g., Moles et al. 1995; Maoiolino et al. 1997; Mulchaey \& Regan 1997). 
The right dynamical and physical conditions must exist for this gas 
to be used effectively in either infalling to the nucleus proper, and feeding the
AGN or nuclear starburst, or collapsing by self-gravity in the circumnuclear 
region in the form of intense SF. There are examples of galaxies where the gas 
is known to be available in the
central regions, but where the onset of SF has yet to occur. For example, 
NGC~4151, which harbours a Seyfert type 1 nucleus, has a circumnuclear 
ring of material discovered in the form of dust extinction, and has a large
column of neutral hydrogen towards the nucleus, but there is no significant 
star formation occurring at present (Robinson et al. 1994; Vila-Vilar\'o et al. 
1995; Mundell et al. 1995). 
However, in the circumnuclear ring of star formation in the target galaxy of
this  paper, NGC~6951, Kohno et al. (1999) find that although the
dynamical shocks provide the mechanism for the accumulation of molecular
gas along the ring, this may not be the mechanism responsible for the 
star formation, but rather the gravitational instability of the clouds thus
formed is the driving force for the onset of star formation.

To understand what are the conditions and mechanisms for the onset
of nuclear activity in spiral galaxies, a detailed characterisation of
morphological and kinematical components in galaxies of different morphology
and activity level is needed. The DEGAS collaboration was established to 
address such a detailed study in a small sample of spirals with an AGN, 
selected as isolated to avoid the possible external contribution to the 
assymmetries in the potential. We plan to procede with a parallel analysis 
of a sample of isolated non-AGN spirals, to look for differences and/or 
similarities and their implications. IR imaging data of the complete sample are 
presented and analysed in M\'arquez et al. (1999a, 1999b). Other recent 
IR imaging studies of active and control samples include McLeod \& Rieke (1995), 
Mulchaey et al. (1997), and Peletier et al. (1998).
 
In this paper we present the first results for the SAB(rs)bc galaxy \object{NGC~6951}. 
There are a number of studies of this galaxy that include 
broad and narrow band optical and infrared images 
(Buta \& Crocker 1993; M\'arquez \& Moles 1993, hereafter MM93; 
Barth et al. 1995; Wozniak et al. 1995; Elmegreen et al. 1996; Friedli et al. 1996; 
Rozas et al. 1996a,b; Gonz\'alez Delgado et al. 1997; Gonz\'alez Delgado \& P\'erez 1997; 
Mulchaey et al. 1997), spectroscopy of the nuclear region  
(e.g., Boer \& Schulz 1993; Filippenko \& Sargent 1985; Mu\~noz-Tu\~n\'on et al. 1989; 
MM93; Ho et al. 1995, 1997b; Ho et al. 1997c), 
radio interferometric maps (Vila et al. 1990; Saikia et al. 1994),
and high resolution spectroscopic molecular maps 
(Kenney et al. 1992; Kohno et al. 1999). 
To better characterize the detailed kinematical components and their relation to the
morphological structures, we have obtained high resolution, high dispersion 
spectroscopic observations of the gaseous and of the stellar components, and direct
imaging in optical and infrared bands.

This paper is structured as follows. In section 2 we present the imaging and
spectroscopic data. Sections 3 and 4 detail the results from the analysis of
the images and from the spectroscopy respectively, while in section 5 we
discuss these results into a common picture of NGC 6951. Section 6 gives the
summary and our main conclusions.


\section{Observations and data reduction}

The journal of observations, given in Table~\ref{tabjournal}, summarizes the
spectroscopic and imaging observations carried out with different telescopes
and instrumental setups, and those retrieved from the HST archives.

\begin{table*}[h!]
\caption{Journal of observations}
\begin{tabular} {lccrlr}
\hline
telescope - PA ({\degr}) & sampling$\rm^a$ & wavelength (\AA) & t$\rm_{exp}$ (s) & airmass  \\
\hline
WHT - 138 & 0.36{\arcsec} & 4757--5141 & 1800 & 1.32   \\
WHT - 138 & 0.36{\arcsec} & 6493--6881 & 1800 & 1.36   \\
WHT - 138 & 0.36{\arcsec} & 8506--8882 & 3600 & 1.34   \\
WHT -  84 & 0.36{\arcsec} & 4757--5141 & 1800 & 1.27   \\
WHT -  84 & 0.36{\arcsec} & 6493--6881 & 1800 & 1.26   \\
WHT -  84 & 0.36{\arcsec} & 8506--8882 & 3600 & 1.26   \\
WHT -  38 & 0.36{\arcsec} & 4757--5141 & 1800 & 1.27   \\
WHT -  38 & 0.36{\arcsec} & 6493--6881 & 1800 & 1.26   \\
WHT -  38 & 0.36{\arcsec} & 8506--8882 & 3600 & 1.26   \\
\hline
NOT       & 0.18{\arcsec} & {\it B\arcmin} & 3000 & 1.36   \\   
NOT       & 0.18{\arcsec} & {\it I}      &  660 & 1.51   \\   
\hline
NOT+ARNICA & 0.51{\arcsec} & {\it J}  & 1760 (440) & 1.25 \\
NOT+ARNICA & 0.51{\arcsec} & {\it Ks} & 3520 (880) & 1.5  \\
\hline
3.5m CAHA  & 0.32{\arcsec} & {\it K\arcmin} &  300 & 1.22   \\
\hline
HST - WFPC2   & 0.046{\arcsec} & {\it V} (547M) &  300 &        \\ 
HST - NICMOS2 & 0.075{\arcsec} & {\it J} (110W) &  128 &        \\
HST - NICMOS2 & 0.075{\arcsec} & {\it H} (160W) &  128 &        \\
\hline
\end{tabular}
\begin{footnotesize}
$\rm^a$ Spatial sampling in arcsec per pixel. The linear dispersion is 
0.39 \AA\ pixel$^{-1}$ for all the spectroscopic observations. 
\end{footnotesize}
\protect\label{tabjournal}
\end{table*}

\subsection{Spectroscopy}

We observed \object{NGC 6951} on the night of 9/10 August 1996 with the ISIS
double spectrograph attached to Cassegrain focus of the 4.2m William
Herschel Telescope, and the gratings B1200 in the blue arm and R1200,
simultaneously, in the red arm. The observations include three wavelength
ranges: one range around H$\beta$ and one around H$\alpha$ (both through the
blue arm of ISIS), and a near infrared wavelength range around the \ion{Ca}{ii}
triplet absorption lines through the red arm of ISIS. At each slit
position angle we took two 1800 s integrations in the red arm, each
simultaneous with one 1800 s in the H$\beta$ and one 1800 s in the H$\alpha$
ranges.

The gratings provide a linear dispersion of 0.39 \AA/pixel, and the
wavelength ranges covered with the TEK1 and TEK2 CCD chips used are
4757-5141 \AA, 6493-6881 \AA\ in the blue arm, and 8506-8882 \AA\ in the red arm. 
We used a slit width of 1 arcsec, that projects onto 2.1 pixels (0.81 \AA) at
the detector. The spatial sampling along the slit is 0.36 arcsec/pixel.
\object{NGC 6951} was observed at three slit position angles: the major axis
at 138\degr, the minor axis at 48\degr, and an intermediate
position at 84\degr.

The observing was performed under photometric atmospheric conditions. At the
WHT the seeing is measured continuously with a guide star; it was slightly 
variable during the observations of \object{NGC 6951}, with a median value of 
$1.2\pm0.1$ arcsec. The phase of the moon was grey.

The reduction of the spectroscopic data followed the standard steps of bias
subtraction, flatfield correction, wavelength calibration with a CuNe lamp
observed before and after the target, atmospheric extinction correction, 
and flux calibration using the spectroscopic standards G24-9, LDS749B, BD284211 
and BD254644 observed through an 8 arcsec wide slit. Cosmic ray hits were
removed individually from the 2D frames. Finally the sky background was
subtracted from each frame using the outermost spatial pixels free from
galaxy contribution. The two exposures in the CaT range were added together
after checking that the spatial and spectral alignment was good to a small
fraction of a pixel. FIGARO and IRAF\footnote{The authors acknowledge the
data analysis software FIGARO provided by the Starlink Project which is run
by CCLRC on behalf of PPARC. IRAF is the Image Analysis and Reduction
Facility made available to the astronomical community by the National
Optical Astronomy Observatories, which are operated by the Association of
Universities for Research in Astronomy (AURA), Inc., under contract with the
U.S. National Science Foundation.} were used throughout the calibration and
analysis phases. 

\subsection{Optical imaging}

Optical images through two continuum filters, {\it B\arcmin} and {\it I}, were taken
with the Brocam II camera on the Cassegrain focus of the 2.56m Nordic Optical
Telescope during the nights of 8 and 9 October 1996. {\it B\arcmin} is a special
purpose made filter with a square spectral response centered at 4630 \AA\
and with a FWHM=290 \AA, so that it does not include any bright emission line.
The equivalent width of any emission line within the {\it I} filter is negligible.
Imaging through emission line free filters is important to ascertain the
true circumnuclear  stellar structure, that can otherwise be significantly
distorted with standard broad band filters, where the emission lines can
make an important flux contribution (Vila-Vilar\'o et al. 1995). The
detector was a 1024$\times$1024 Tektronix chip with a sampling of 0.176
arcsec/pixel and a 3$\times$3 arcmin field of view. Weather conditions were
photometric and the average FWHM of the seeing profile is 0.8 arcsec. The
journal of observations is included in Table~\ref{tabjournal}.

The reduction and calibration of the images has been carried out in IRAF.
Bias and flat field corrections were done in a standard way. The task
COSMICRAYS was used for the overall cosmic ray removal.

The final image of \object{NGC 6951} for a given filter is obtained by
adding all the images for that filter properly aligned, sky subtracted and 
calibrated. The alignment, using cross-correlation techniques involving
only bright stars, is made to an accuracy of 0.1 pixel. The sky background
is estimated by averaging the median flux in a few 40$\times$40 pixel
patches located outside the galactic disk. 
The 3$\sigma$ background limit magnitudes are 20.7 and 21.0 mag~arcsec$^{-2}$ 
in {\it B\arcmin} and {\it I}, respectively.

\subsection{Infrared imaging}

Infrared images through the filter {\it K\arcmin} ($\lambda_c$=2.10 $\mu$, 
width=0.34 $\mu$) were taken with the MAGIC camera attached to the
Cassegrain focus of the 3.5m telescope at the German-Spanish Observatory in
Calar Alto on 1996 September 26. The detector was a Rockwell 256$\times256$
pixel NICMOS3 array, and the camera was set up in the high resolution mode
of the f/10 focus, giving a sampling of 0.32 arcsec/pixel and a field of
view of 82\arcsec$\times$82\arcsec. The observation procedure consisted in
3$\times3$ mosaic patterns of the object and nearby sky, alternating the exposures. 
The mosaic on the galaxy has a 90\% overlap between the component tiles. During 
the observations, the seeing was 1.0 arcsec and the conditions photometric.
Each galaxy frame was subtracted from a median sky frame and divided by the flat
field. The nine frames were recentered using two or three field point sources 
and the nucleus, and subsequently they were median averaged. The flux calibration 
was performed with observations of the Kitt Peak faint standard stars numbers 
30, 31 and 32.  
  
NGC~6951 was observed on 1996 September 30, as one of the targets of the BARS 
collaboration, through the filters {\it J} and {\it Kshort} ({\it Ks}) with 
the ARcetry Near-Infrared 
CAmera, ARNICA, attached to the 2.56m Nordic Optical Telescope in La Palma. 
ARNICA uses a NICMOS3 array that gives a pixel scale of 0.51 arcsec/pixel and 
a field of view of 130\arcsec$\times$130\arcsec. We integrated 40 seconds on 
the object (dithered in 11 positions) and the same time on the sky, for a total
integration time on the object of 440 s in {\it J} and 880 s in {\it Ks} (with a 
{10\arcsec} shift between the different images to facilitate bad pixel removal). 
Since the size of the galaxy is much larger than the field of view, we repeated
the process four times, with the central region of \object{NGC 6951} in common,
placed in one of the corners. This allowed us to obtain a final mosaic image
with a total integration time of 1760 s in {\it J} (3520 s in {\it Ks}) in the 
central {2\arcmin} 
and 440 s in {\it J} (880 s in {\it Ks}) in the rest. Flat-field frames were obtained from 
the median of the sky frames taken along the night. Sky-subtracted and flat-fielded 
images were then aligned by using a number of foreground stars. We observed 4
standard stars for the flux calibration, leading to a
photometric accuracy of 10\%. For all the reduction and calibration steps we
used the package {\tt sqiid} within IRAF. We measure a FWHM for the stars
in the final mosaic of 1.3\arcsec. The 3$\sigma$ limiting magnitudes are 18.8 in 
{\it J} and 19.5 in {\it Ks}.
 
\subsection{HST images}

We have retrieved archival HST images taken with the WFPC2 through the {\it V} 
filter (547M), with a single exposure of 300 s, and with the NICMOS2 camera 
through the {\it J} (110W) and {\it H} (160W) filters, with exposures of 128 s each. 
The spatial 
sampling of the central PC image in the WFPC2 camera is 0.046{\arcsec}, and  
0.075{\arcsec} for the NIC2 camera of NICMOS. The useful field of view is 
$\sim$12{\arcsec}$\times$12{\arcsec}, which contains the circumnuclear region.

 
\section{Imaging results}  
 
Fig.~\ref{imBK} shows the {\it B\arcmin} and Ks images of the galaxy.  
The most conspicuous features in both images are the bar,
the bulge and the spiral arms. The bar is better traced in {\it Ks}, 
since the distortions produced by the dust lanes are minimized. 
The circumnuclear ring and dust lanes are more clearly 
traced in {\it B\arcmin}, where they appear as straight features from the bar edges 
towards the center, curving approximately  within the bulge radius. All the ring 
features reported by Buta \& Crocker (1993) are visible: (pseudo) outer, 
(pseudo) inner and circumnuclear rings, with sizes in agreement with those given 
by them, and also with optical and/or near-infrared images previously published 
(MM93, Friedli et al. 1996). We note the isophote twisting in the 
central parts, from the  bar to the circumnuclear ring. 

To parametrize the results of the photometric analysis, we have obtained the
isophotal profiles by using the task {\tt ellipse} within
{\tt stsdas.analysis.isophote} in IRAF, which uses the ellipse fitting method.
All the stars in the frame were previously masked.  The results are given in
Fig.~\ref{parBIJK}, where we plot isophotal magnitudes, ellipticity and
position angle (PA) as a function of the square root of the radial distance
for {\it B\arcmin}, {\it I}, {\it J} and {\it Ks} bands. They agree with previous 
results by Friedli et al. (1996), who observed NGC~6951 with a similar configuration,
although our data extends 30 arcsec further out. Since the area covered by our
optical images is smaller than for MM93, and considering that the faintest
isophotes we reach in NIR are still contaminated by the spiral arms, we have
decided to use the disk major axis  position angle and inclination as
determined by MM93, i.e., $\phi=138${\degr} and $i=42${\degr}. These values
are compatible with the outermost ellipses as traced by the sharp-divided
images, and are in good agreement with the kinematically derived position
angle and inclination, as we will see below.   

To quantify the bar extension we have computed the amplitude and phase of the
Fourier transform applied to the deprojected  image, as explained in M\'arquez
\& Moles (1996). The mode m=2 has a constant phase out to
45{\arcsec}, that we adopt as  the bar extension.

The isophotal profiles indicate the presence of the circumnuclear ring and bar
components, that appear as an excess around 5{\arcsec} for the former and from 20{\arcsec}
to 50{\arcsec} for the second. These excesses are clearer once the 
photometric decomposition is made. We have fitted an exponential disk for the 
outermost radii, subtracted it from the profile, fitted an r$^{1/4}$ for the 
resulting bulge, subtracted it from the profile, and iterated until convergence 
is achieved. The results from the decomposition should be taken with caution since
the bar contamination and the dust obscuration are very important. The results for the
infrared image are presented in M\'arquez et al. (1999).

{\it B{\arcmin}-I} and {\it J-Ks} full size colour images are shown in Fig.~\ref{NOTBINOTJK}.  
In {\it B{\arcmin}-I} the reddest features trace the dust lanes, most outstanding in the 
western side of the bar, while the bluest colours trace the regions of active
star formation, that occur in the spiral arms outside a 30 arcsec radius. 
Within the part of the galaxy dominated by the bar structure there is very little 
ongoing star formation, except in the circumnuclear ring that is delineated by a
number of blue knots. A comparison with the H$\alpha$ image (fig. 5 in MM93; fig. 10
in Gonz\'alez Delgado et al. 1997) shows that, for most regions, the
H$\alpha$ flux correlates with the colour {\it B{\arcmin}-I}, however the bluest large
region at 46 arcsec along position angle 20\degr\ does not have a
correspondingly bright H$\alpha$ emission, indicating that it may be a star
forming region in the post-nebular phase. A detailed analysis of the stellar
populations in NGC~6951 will be presented in a future paper. In the two colour maps,
bulge and bar colours are very similar.

The highest contrast in the {\it J-Ks} image occurs around the western side of the 
circumnuclear ring, and along the western dust lane close to the ring. To look at
these in more detail, Fig.~\ref{BKIK} shows {\it B{\arcmin}-K{\arcmin}} and {\it I-K{\arcmin}}
colour maps, where {\it K\arcmin} is the Calar Alto image, that has better sampling and seeing.
Along the dust lanes, the colour contrast is higher in {\it B{\arcmin}-K{\arcmin}}, that
results from a larger dust absorption in {\it B\arcmin} than in {\it I}. For the reddest
regions inside the circumnuclear ring, the situation is the opposite, with redder
colours in {\it I-K{\arcmin}}, in particular, close to the two ends of the ring major axis 
(NW-W and SE-S), and the SW zone where the ring connects with the bar. This
higher contrast in {\it I-K{\arcmin}} is similar to {\it J-Ks} in Fig.~\ref{NOTBINOTJK}. 
The smaller contrast of the dust in {\it B{\arcmin}-K{\arcmin}} can be explained by an 
excess of {\it B\arcmin} produced by the ongoing star formation in the ring. Thus the
colour maps can be qualitatively understood in terms of the combined effects of 
dust obscuration and stellar population.

The HST WFPC2 {\it V} image has been published and analysed by Barth et al. (1995), that
study the distribution of super star clusters in the circumnuclear ring. 
The HST NICMOS images have not been published yet. In Fig.~\ref{HST1} we show the 
{\it J-H} colour image. The red features contouring the circumnuclear ring are visible 
in great detail, mainly delineated as two short spiral arms that trace the 
ends of the oval. Closer to the nucleus, it consists of a number of small knots
and structures that delineate a multi-spiral like structure.


\section{Spectroscopic results} 

\subsection{The emission line spectrum}

The three slit orientations used are depicted in Fig.~\ref{BKHanucleo}, 
that shows the H$\alpha$ contours (from the data of Gonz\'alez Delgado et al. 1997)
on a B{\arcmin}/K{\arcmin} grayscale colour map. The three position angles 
correspond to the major axis (PA=138{\degr}), the minor axis (PA=48{\degr}), 
and an intermediate angle (PA=84{\degr}). The slits cross the circumnuclear 
ring, but only one of the bright H$\alpha$ knots.

We have measured the emission line intensities at every spatial increment 
along the slit, by means of fitting gaussian components with LONGSLIT 
(Wilkins \& Axon 1991). Fig.~\ref{hahb} shows the spatial variations 
of the H$\alpha$ line intensity and the H$\alpha$/H$\beta$ ratio for 
position angles (a) 48\degr, (b) 84{\degr} and (c) the major axis at 138{\degr}. 
The Balmer lines are affected by underlying absorption
components in the nucleus and in the ring \ion{H}{ii} regions. A detailed
analysis of the stellar population contributing to the Balmer absorption is
beyond the scope of this paper, but a comparison of the spectra with 
evolutionary population synthesis profiles of the Balmer absorption line
profiles (Gonz\'alez Delgado et al. 1999b), indicates that the Balmer 
emission lines should be corrected by a line core equivalent width of 1 \AA\ in 
the nucleus and 0.5 \AA\ in the ring; this result is found through the 
comparison of the absorption wings detected in the H$\alpha$ and H$\beta$
lines and in the nearby metal absorptions with the predictions of the models.  
The H$\alpha$/H$\beta$ ratio shown in Fig.~\ref{hahb} has been corrected in
this manner. The Balmer emission line ratio maps the effect of extinction by 
the dust lanes; it is high in the nucleus (H$\alpha$/H$\beta\geq10$ at the center), 
then becomes smaller in the transition region between the nucleus and the ring 
of \ion{H}{ii} regions (H$\alpha$/H$\beta\simeq4$ at a distance of $\pm$1.5 arcsec), 
and increases again across the ring to reach values of H$\alpha$/H$\beta\simeq10$ 
in the outer side of the ring. This reddening across the ring is
clearly produced by the dust lanes, as seen in Fig.~\ref{BKIK}. 
The extinction that corresponds to the above values of the
Balmer ratio is c(H$\beta$)=1.6, 0.4, and 1.6 mag for the nucleus, the
transition region and the ring respectively. Thus, just outside the partially
resolved nucleus, at $\pm$1.5 arcsec, the extinction reaches a minimum and the
Balmer ratio takes the case B value. The HST image in Fig.~\ref{HST1} shows
indeed little dust in that intermediate region. The Galactic extinction towards
\object{NGC 6951} quoted in the NED IPAC database is 0.88 mag in {\it B}, corresponding
to E({\it B-V})=0.22 or c(H$\beta$)=0.31, i.e. most of the extinction that we
measure in the intermediate region between the nucleus and the circumnuclear ring 
is of Galactic origin and not intrinsic to NGC~6951.
 
Fig.~\ref{n2ha} shows the line ratio [\ion{N}{ii}]6583/H$\alpha$ together 
with the H$\alpha$ and [\ion{N}{ii}] fluxes. For all three PA the ratio becomes 
larger than unity within the inner $\pm$2 arcsec, reaching values of 5 in the 
nucleus. This large ratio is due to a strong increase in the [\ion{N}{ii}] flux
and not to underlying absorption in H$\alpha$. Indeed, the [\ion{N}{ii}] takes 
normal \ion{H}{ii} region values in the ring, but increases sharply inwards. 
This region of high [\ion{N}{ii}]6583/H$\alpha$ 
is significantly more extended than the nuclear point spread function, and
thus  it is spatially resolved. The [\ion{S}{ii}]/H$\alpha$ ratio (not
shown) also shows a qualitatively  similar behaviour, with an
increase from [\ion{S}{ii}]/H$\alpha$=0.18 in the ring \ion{H}{ii}  regions to
[\ion{S}{ii}]/H$\alpha\geq1.8$ in the nucleus. The electron density, computed 
from the line ratio [\ion{S}{ii}]6717/6731, increases from low values,
$N_e\sim300~\rm cm^{-3}$,  in the ring \ion{H}{ii} regions to values larger
than 1000 cm$^{-3}$ within  the inner region and the nucleus (cf. Table
\ref{tabeden}). These [\ion{N}{ii}]  and [\ion{S}{ii}] line ratios seem to
indicate the presence of a shocked component,  also supported by kinematical
evidence, as explained below.

\begin{table*}[h!]
\caption{Electron density along the slit}
\begin{tabular} {lrr}
\hline
PA & distance & density   \\
   & (arcsec) & cm$^{-3}$ \\
\hline
138{\degr} SE  & -4.5 &  530 \\
138{\degr} SE  & -1.1 & 1440 \\
138{\degr} nuc &  0.0 & 1340 \\
138{\degr} NW  &  1.0 &  990 \\
138{\degr} NW  &  4.5 &  290 \\
\hline
\end{tabular}
%
\protect\label{tabeden}
\end{table*}

Fig.~\ref{hbo3} shows the H$\beta$ and [\ion{O}{iii}] 5007 line fluxes along 
PA 138\degr. The nucleus has a high excitation. The [\ion{O}{iii}] flux 
decreases sharply to reach a ratio [\ion{O}{iii}]5007/H$\beta<1$ outside 
$\pm2$ arcsec from the nucleus. Thus the ring \ion{H}{ii} regions have very 
low excitation; in fact, we cannot measure 
[\ion{O}{iii}] 5007 pixel by pixel outside the nuclear component at any of the 
three position angles. In order to compute this excitation ratio for the 
ring \ion{H}{ii} regions, we have extracted three one-dimensional spectra at 
each position angle, integrating the nucleus and the two sides where the slit 
crosses the ring. Fig.~\ref{hbspec} shows these for PA=48\degr. 
The measured H$\beta$ and [\ion{O}{iii}] 5007 line fluxes and the ratio 
[\ion{O}{iii}]5007/H$\beta$ are given in Table \ref{tabhbo3}, as well as 
the ratio with the H$\beta$ flux corrected for underlying absorption 
(an absorption equivalent width of 1 \AA\ in the nucleus and 0.5 \AA\ 
in the \ion{H}{ii} regions).

\begin{table*}[h!]
\caption[ ]{High excitation ratio [\ion{O}{iii}]/H$\beta$}
\begin{tabular} {rlccrc}
\hline
PA & orientation & extraction & F(H$\beta$) & F(5007)/F(H$\beta$) & F(5007)/F(H$\beta$) \\
   &             &   arcsec   & $\rm10^{-15}~erg~s^{-1}~cm^{-2}$ & &    corrected$^a$ \\
\hline
138{\degr} & SE  & -5.69,-2.82 & 1.13 &  0.08 & 0.07 \\
138{\degr} & nuc & -1.40, 1.11 & 0.20 & 16.26 & 5.51 \\
138{\degr} & NW  &  2.54, 5.40 & 1.17 &  0.03 & 0.03 \\
\hline
 84{\degr} & E   & -4.00,-1.80 & 0.38 &  0.36 & 0.29 \\
 84{\degr} & nuc & -1.10, 1.40 & 0.17 & 24.82 & 7.03 \\
 84{\degr} & W   &  2.70, 4.90 & 0.52 &  0.14 & 0.12 \\
\hline
 48{\degr} & NE  & -4.11,-1.61 & 0.56 &  0.30 & 0.25 \\
 48{\degr} & nuc & -0.89, 1.25 & 0.37 & 16.57 & 7.96 \\
 48{\degr} & SW  &  1.97, 4.47 & 0.23 &  0.53 & 0.39 \\
\hline
\end{tabular}

\begin{footnotesize}
$^a$ H$\beta$ corrected for an absorption equivalent width of 1 \AA\ in the 
nucleus and 0.5 \AA\ in the \ion{H}{ii} regions.
\end{footnotesize}

\protect\label{tabhbo3}
\end{table*}

\subsection{The absorption calcium triplet spectrum}

We have measured the equivalent width of the two main CaT lines at 
$\lambda \lambda8542,8662$\AA, according to the method described by D\'\i az, 
Terlevich  \& Terlevich (1989, hereafter DTT). The nuclear spectrum 
(central 0.72 arcsec) is plotted in Fig.~\ref{CaTspec} for reference; 
the three CaT lines, MgI $\lambda8807$, and two of the hydrogen Paschen 
lines in absorption are identified. The weakness of Pa14 indicates
a negligible contamination of the CaT lines by Pa16, Pa15 and Pa13, that fall 
in their red wings. The results of the pixel to pixel 
equivalent width measurements along the slit, for the three position angles, 
have been combined and are shown in a single Fig.~\ref{ewCaT}. The lefthand 
side of this figure, (-10 arcsec to the nucleus) corresponds to the NE--SE 
quadrant on the eastern side of the galaxy. The righthand side of the figure 
(from the nucleus to 10 arcsec) corresponds to the NW--SW quadrant on the 
western side. There is a clear systematic trend in this combined figure, 
that can also be appreciated (albeit with a worse signal-to-noise ratio) 
along the three individual PA. Outside the circumnuclear  ring, ew(CaT) 
takes a low value of $\sim5.5$ \AA; at and within the ring, 5 arcsec east 
and 4 arcsec west, ew(CaT) jumps to a higher value of 7 \AA; and closer in, 
within the partially resolved nucleus, ew(CaT) rises to a value 
between 8 and 9 \AA. These are three distinct regimes clearly present in the 
circumnuclear region.

The interpretation of ew(CaT) as a function of a stellar population depends 
on the metallicity (DTT; see the most recent work by Garc\'\i a-Vargas, 
Moll\'a  \& Bressan, 1998, and references therein). We can have an indication 
of the metallicity by measuring the ew(MgI). According to DTT this line is 
sensitive to metallicity and to effective temperature, but not to gravity. 
In the nucleus of NGC~6951, ew(MgI)$\geq0.8$ \AA, that indicates 
(see fig. 8 of DTT) the production by cool ($\rm T_{eff}<4700$ K), 
high metallicity stars (solar or higher). For this metallicity range, 
ew(CaT) is a function mainly of gravity and so we can now interpret the 
measurements in terms of stellar populations.

The lowest values of ew(CaT), present mainly outside the ring, 
are due to an old stellar population of giants, with an age around 0.5 to 1 Gyr. 
The highest nuclear values, ew(CaT)$\geq8$ \AA\ indicate the presence of an 
important population of red supergiant stars (RSG), that dominate the 
luminosity at these wavelengths, with an age around 10 to 20 Myr. 
In the circumnuclear  region, ew(CaT) takes intermediate values; 
it is significantly higher than in the outer parts, but the actual 
value belongs to a range that can be explained by either giants or RSG. 
Our interpretation is as follows. The measured equivalent widths in the 
circumnuclear  and nuclear regions are a lower limit to the actual values, 
for two reasons. First, these equivalent widths are diluted by the 
continuum of the young stellar population responsible for the ionization 
of the gas; we cannot precisely gauge with our present data set how important 
is this dilution, for which we would need spectroscopy in a longer 
wavelength range, including the ultraviolet, but we do see the blue stellar 
knots so that some dilution by them must be taking effect. Second, at least 
outside a radius of 2 arcsec, we can distinguish two distinct stellar 
kinematical components; although we cannot measure independently their 
contribution to the luminosity or to the ew(CaT), it is at least possible 
that one of these two distinct  populations (one that can be 
identified with the bar population) contributes to dilute the other population 
(whose kinematics is dominated by rotation, see below). Thus, if dilution is 
taken into account it is likely that the circumnuclear  values of 
ew(CaT)$\sim7$ \AA\ are in fact larger, so that they also reflect a
population  of RSG within the ring.

\subsection{Gas kinematics}

We have measured the systemic velocity in H$\alpha$ by integrating all 
the spatial increments along the slit corresponding to the disk emission
(i.e. excluding the nucleus and the circumnuclear ring of \ion{H}{ii}
regions), and fitting the two peaks of the resulting integrated profile. The
mean  velocity of these two peaks is 1417$\pm$4 km s$^{-1}$; this corresponds 
to a scale of 92 pc/arcsec, for H$_0$=75 km s$^{-1}$ Mpc$^{-1}$. The velocity 
curves obtained from the pixel to pixel measurements of the H$\alpha$ 
emission line are shown in Fig.~\ref{veloflux12}; the H$\alpha$ line flux 
is also plotted for reference. Allowing for the different spatial and 
spectral resolutions, these velocity curves agree very well with those 
presented by MM93. In the following we analyze the detailed and the global 
kinematic structures found along the three position angles.

First we confirm that PA=138{\degr} presents the largest velocity amplitude, 
followed by PA=84{\degr} and PA=48\degr, indicating that 
138{\degr} is closest to the major axis. At first glance, there seems 
to be much velocity structure at PA=48{\degr} that does not correspond 
to a kinematic minor axis, where we would not expect a systematic residual 
velocity curve. On a more detailed analysis, this velocity structure present 
along PA=48{\degr} is related to local morphological structure, 
and we can see that the outer parts of the disk velocity correspond to the 
systemic velocity, as indicated by a horizontal dotted line, 1417 km~s$^{-1}$. 
Thus we adopt 48{\degr} and 138{\degr} as the kinematic minor and 
major axes respectively.

For this value of the kinematic major axis and an inclination angle of 
42{\degr} (MM93), the total dynamical mass enclosed 
within the circumnuclear region (a radius of 6 arcsec, equivalent to 644 pc), 
is $M=r\,v^2/G=5.8\times10^9$ M$_{\odot}$. Kohno et al. (1999) report a 
\emph{molecular} gas mass of $1.4\times10^9$ M$_{\odot}$ within a 6 arcsec radius;
this amounts to 25\% of the total dynamical mass.

Much detailed structure is present in the velocity curves at the different PA. 
For example, at PA=48{\degr} we detect the 50 {km~s$^{-1}${} streaming into 
the main NE arm, between -56.5 arcsec and -53 arcsec (labelled A in
Fig.~\ref{veloflux12}).  Also at this PA we detect the 50 {km~s$^{-1}${}
streaming into the bar region  at -36 to -20 arcsec in the NE (labelled B),
and at 32 to 23 arcsec to the SW  (labelled B{\arcmin}). At PA=84\degr, the 50
{km~s$^{-1}${} streaming  into the bar is seen at the eastern edge, between 40
and 30 arcsec (labelled C),  and at the western edge of the bar between 39 and
32 arcsec (labelled C{\arcmin}).

When deprojected for an inclination of 42{\degr} and a kinematic major 
axis of 138\degr, the velocity curves at PA 84{\degr} and at 
PA 138{\degr} generally match in the outer disk regions, except for 
local distorsions produced by the spiral arms. However, the circumnuclear  
region shows a different picture. One important feature of the velocity 
curve at PA 48{\degr} is the existence of an {\em apparent} counter-rotation 
of the gas within the circumnuclear  region with respect to the gas in the 
disk at the same PA. Indeed, when we look into the $\pm$8 arcsec circumnuclear  
velocity gradients at the three position angles more closely (insets in 
Fig.~\ref{veloflux12}), we see three features. First, if we deproject the
circumnuclear  velocity curves using the same two parameters as for the 
main disk, $i=42${\degr} and $\phi=138${\degr}, the deprojected amplitude at 
PA=84{\degr} becomes significantly larger than at PA=138\degr; 
this would imply that the circumnuclear  region requires a different set 
of deprojection angles. Second, there is a continuous change of slope 
between the three PA, flattening from PA=138{\degr} to PA=84{\degr} 
to PA=48\degr. If the circumnuclear  velocity curve along PA=48{\degr} 
is also produced by disk rotation, then the kinematic axis of this 
circumnuclear  rotation must be different from the kinematic axis of 
the main disk. Under this assumption of disk rotation, we can compute a 
kinematic major axis for this circumnuclear region. We obtain that in 
this case the major axis of the circumnuclear region would correspond 
to 118\degr; with this circumnuclear major axis and for the 
same\footnote{If this circumnuclear region is a decoupled gaseous disk, 
it could have a different inclination to the line of sight that the main 
galaxy disk, but we will assume here that this is not the case and that 
the inclination is the same.} inclination of 42\degr, the three 
deprojected circumnuclear  velocity curves agree quite well. Third, 
at PA=84\degr, the circumnuclear  velocity amplitude, 200 {km~s$^{-1}${}, 
is significantly larger than the galaxy main disk velocity amplitude at 
the same PA, 130 {km~s$^{-1}${}. These three facts, namely, the {\em apparent}
nuclear counter-rotation at PA=48{\degr} and the greater velocity 
amplitude at PA=84{\degr} (both, with respect to the disk amplitude 
at PA=84\degr, and with respect to the deprojected circumnuclear  
amplitud at PA=138\degr), argue for a gas dynamics in the circumnuclear  
region of NGC~6951 that is decoupled from the dynamics of the main body 
of the galaxy. We shall return to this point in the discussion section.

The FWHM of H$\alpha$ and [\ion{N}{ii}] are plotted in Fig.~\ref{fwhm}. 
The values shown have been corrected for an instrumental resolution of 43.5 
{km~s$^{-1}${}. The influence of the nuclear velocity gradient on the widths 
is negligible; the gradient is 44 {km~s$^{-1}${} arcsec$^{-1}$, that amounts 
to a quadratic correction on the nuclear widths of only 2 {km~s$^{-1}${}.
The H$\alpha$ line flux is shown (dotted) for reference. The velocity dispersion 
in the circumnuclear region is very high, in the H$\alpha$ line there is a 
central plateau of 180 {km~s$^{-1}${} in the nuclear $\pm$1.2 arcsec, 
it then decreases down to a minimum of 70 {km~s$^{-1}${} 
that is reached at the maximum of H$\alpha$ emission in the \ion{H}{ii} regions. 
After this local minimum, the velocity dispersion increases again to 120 
{km~s$^{-1}${} and above in the two quadrants NE-NW and SE-SW across the ring 
of \ion{H}{ii} regions. This trend is consistent with the general picture of 
the dynamics deduced from our data and from the molecular content 
(Kohno et al. 1999), that will be developed in the discussion section. 
The behaviour of the velocity dispersion for the [\ion{N}{ii}] $\lambda$6583 
line is qualitatively similar to but systematically higher than that of 
H$\alpha$ by a factor 1.2 in the inner $\pm$3 arcsec, reaching values of up to 
220 {km~s$^{-1}${} in the nucleus. These emission line widths indicate 
dynamically hot nuclear and circumnuclear regions. The kinematic data and the 
results from the emission line ratios and density values presented in section 
4.1 seem to imply the existence of shocked material. 

The width of H$\alpha$ is not resolved further out in the disk of the galaxy, 
except across the spiral arms where it reaches values of 25 to 45 {km~s$^{-1}${}.

\subsection{Stellar kinematics}

We have measured the velocity curve from the two main lines of the \ion{Ca}{ii}
absorption triplet at $\lambda\lambda8542,8662$ \AA,  
by means of cross-correlating the galaxy frames with seven spectra of five 
different K giant stars, observed during the same night and with the same setup. 
These stars are HD132737, HD171232, HD208817, HD198858, and HD1918. 
The cross-correlation is performed with a 2D procedure we have implemented that
makes use of the IRAF function CROSSCORR. The position and amplitude of the peak 
of the cross-correlation function is measured, and the resulting velocity curves
are weight averaged at each position angle. The signal-to-noise ratio of the 
data allows measurements only within a 20 arcsec radius at PA=84{\degr} and 
48{\degr}, and 10 arcsec at PA=138\degr. The resulting curves are shown in 
Fig.~\ref{veloCaT12} (expanded for the central region in the insets) 
for the three position angles, together with the corresponding H$\alpha$ emission 
velocity curves for comparison. 

First, we inspect the stellar velocity curves. Inwards of the circumnuclear ring, 
in the $\pm$2 arcsec radius resolved nuclear zone, 
there seems to be a normal stellar rotation curve, with a measured gradient 
of 41 and 26 {km~s$^{-1}${} arcsec$^{-1}$ at PA=138{\degr} and 84\degr, 
and flat at PA=48\degr. Outside this radius it is possible to partially
resolve  the absorption line profile into two components, that are seen in the 
cross-correlation function by corresponding local peaks; in most spatial 
increments one component fully dominates the profile, 
while in some increments the position of the two components can be more easily 
measured. The velocities corresponding to these two components are plotted with 
filled and open symbols in the figure. On either side of the nucleus, 
these two stellar components have their velocity of the same sign, although 
their amplitude difference reaches 50 km~s$^{-1}$. 

We now look into the comparison with the gas H$\alpha$ velocity curves. Where only 
one stellar component is measured, i.e. within the nuclear 2 arcsec radius, the 
velocity curves of the gas and of the stars at PA=84{\degr} and 138{\degr}
follow each  other approximately. This is not the case at PA=48\degr, 
where the stellar velocity curve is flat, while the H$\alpha$ curve presents 
a slope (c.f. previous section 4.3) with a peculiar structure in the central 
$\pm$1.5 arcsec. 

At PA=84{\degr} the nuclear $\pm1.5$ arcsec part of the gaseous and stellar 
velocity curves both follow the same similar pattern; however, when the two 
stellar components can be kinematically resolved, neither of them two match 
the gas kinematics. One component is very flat with a mean velocity of 1440 
{km~s$^{-1}${} from 4.5 arcsec eastwards of the nucleus, and a velocity of 1400 
{km~s$^{-1}${} from 5 arcsec westwards of the nucleus. This stellar component 
matches the velocity of the gas entrained in the bar both at 30 arcsec east 
(at 1440 {km~s$^{-1}$) and at 30 arcsec west (at 1400 {km~s$^{-1}$; c.f. previous 
section 4.2 and Fig.~\ref{veloCaT12}). We shall refer to this component as the 
\emph{stellar bar component}. The other stellar component is steeper but not 
as much as the gas component between $\pm$3 arcsec and $\pm$7 arcsec, where the 
gas presents a significantly larger rotation amplitude than the stellar disk 
rotation at the same projected distance (a similar case to that reported for the 
barred galaxy NGC~6701 by M\'arquez et al. 1996, cf. their fig. 12). The extrapolation 
of this stellar component meets the gas rotation in the disk at 40 arcsec east 
(at 1490 {km~s$^{-1}$) and at 40 arcsec west (at 1330 {km~s$^{-1}$); we shall 
refer to this component as the \emph{stellar disk component}.

At  PA=138{\degr} the stellar velocity presents a similar behaviour. 
Only one component is apparent within a 3 arcsec radius; this generally 
follows the gas velocity, although with less structure in the nucleus. 
Outwards of this radius two stellar velocity components can be distinguished. 
At the south-east the steeper of the two components reaches a maximum velocity 
of ~1553 {km~s$^{-1}$, that extrapolates to match the disk gas velocity. 
 
The stars and the gas kinematics are most different at PA=48{\degr}.
The gas presents an {\em apparent} counter-rotating component within a 6 arcsec 
radius with respect to its outer disk rotation, and an additional peculiar 
velocity structure in the nuclear $\pm$1.5 arcsec. The stars show a slowly 
rising kinematics from north-east to south-west, compatible with the disk 
rotation, and there is no indication of any peculiar velocity structure in the 
nucleus similar to the gaseous one. This peculiar structure in the gas velocity
along the minor axis could be due to an off-centered slit, to a gaseous nuclear
outflow, or to a structure related to the iILR.


\section{Discussion}

We shall now discuss the results obtained in the previous sections and try to
put them into a coherent picture of the morphology and kinematics of the circumnuclear
region in NGC~6951.

\subsection{The level of activity in the nucleus: Seyfert 2 or LINER?}

The active nucleus in NGC~6951 has been classified in the literature as a
Seyfert 2 (e.g., Boer \& Schulz 1993; Ho et al. 1995, 1997b) or as a LINER 
(Filippenko \& Sargent 1985; Mu\~noz-Tu\~n\'on et al. 1989; 
MM93). Of the different classifications, that of Ho et al.
(1997b) is the most accurately performed, because of the data quality  
and the correction for underlying absorption, that affects 
somewhat to H$\alpha$, and more to H$\beta$. The classification is based on 
the three Veilleux \& Osterbrock (1987) diagrams
[\ion{O}{iii}]5007/H$\beta$ vs. [\ion{N}{ii}]6583/H$\alpha$,
[\ion{O}{iii}]5007/H$\beta$ vs. [\ion{S}{ii}]6716+6731/H$\alpha$, and 
[\ion{O}{iii}]5007/H$\beta$ vs. [\ion{O}{i}]6300/H$\alpha$. 

We have extracted the central $1.0\times1.2$ arcsec as the most representative
spectrum of the nucleus, and we measure [\ion{O}{iii}]5007/H$\beta$=2.59 (5.89),
[\ion{N}{ii}]6583/H$\alpha$=7.00 (3.72), and [\ion{S}{ii}]6716+6731/H$\alpha$=3.06 
(1.63). For the values in parentheses, H$\beta$ and H$\alpha$ have been
corrected for an underlying 1 \AA\ of equivalent width in absorption (as discussed 
in section 4.1 above); these are the values used in the diagrams. Further, 
recently we have had access to a lower resolution spectrum, from where we can measure 
[\ion{O}{i}]6300/H$\alpha\geq$0.35. Using these values in the diagrams, 
the nucleus of NGC~6951 is located in the very high excitation end of the region
of these diagrams filled by the LINER points. This explains the uncertain 
classification, even when the correction for underlying absorption is taken into 
account. In fact, because the most critical value is [\ion{O}{iii}]5007/H$\beta$
which is most sensitive to this absorption correction, it makes the precise
definite classification somewhat arbitrary. Ho et al. (1997b) classification
criteria (cf. their table 5) consider a horizontal boundary between Seyfert 2 and 
LINERs at [\ion{O}{iii}]5007/H$\beta\geq$3; this puts NGC~6951 in their Seyfert 2
zone. Other authors consider that the boundary between the two classes 
changes with the value of the abscissa (e.g. Gon\c calves et al. 1999), 
and this would put NGC~6951 in the higher excitation end of the LINER zone. 
In either case, the value of [\ion{N}{ii}]6583/H$\alpha$=3.72 is also very high, 
that may suggest an overabundance of nitrogen; this will be discussed in a future 
work on the stellar populations and gas phase abundances.

In summary, the active nucleus in NGC6951 can be considered as a transition object
between a very high excitation LINER and a possible nitrogen overabundant Seyfert 2.

\subsection{Colour maps: dust and stellar populations}

According to numerical simulations and to observations of barred galaxies,
the interaction of the bar and the disk creates mildly shocked zones
that enhance the formation of \ion{H}{ii} regions, mainly at the ends and along the
leading edges of the bar. This also produces a net flow of disk gas and dust 
towards the central regions of the galaxy, where the
interaction with the bulge and nuclear dynamics often enhances the 
circumnuclear  star formation later in the evolution of these systems. 
In Fig.~\ref{NOTBINOTJK}, the scarcity of star forming regions in the bar 
(both, at the ends and along the leading front of the bar), and the presence 
of the circumnuclear ring of star formation, argues for a relatively old 
dynamical age of the bar and associated systems in NGC~6951. Other imaging 
and spectroscopic results give further support to this scenario.
 
When we consider all the information available (cf. sections 3 and 4), 
a pattern seems to emerge in the circumnuclear region in which: \\
(a) There is a widespread population of supergiant stars that dominate the 
light along and inwards of the circumnuclear  ring, between about 6 arcsec
and 1.5 arcsec radius. These are most clearly seen in the {\it K} dominated 
{\it J-K} map,
together with the large CaT equivalent widths measured in this same
zone (cf. section 4.2). \\ 
(b) The correspondence between the regions more luminous in 
{\it B\arcmin} and in
H$\alpha$ is only good for the two brightest ones, but there are two regions 
bright in H$\alpha$ with apparently no corresponding enhanced {\it B\arcmin} that must be
understood in terms of extinction; and there is widespread enhanced B{\arcmin} along
and inwards of the ring with little or no corresponding H$\alpha$ emission, 
that indicates star formation in the post-nebular phase. It is interesting to
notice that {\it B\arcmin} is enhanced both locally in a knotty structure along the ring
and more diffuse along and inwards of the ring; this might be interpreted as a 
signature of two modes of star formation coexisting here: continuous and
bursting. It is possible that a continuous star formation process with 
ocasional bursts takes place in a region such as this one, where a more 
or less steady supply of gas has been going on for a long time, as supported by the 
large fraction of molecular to total mass within the region (cf. section 4.3 above). \\
(c) There is some indication of a time sequence in the star formation history
along the ring, more apparent in the W-N quadrant. First, the strong {\it K} knot at 
position angle $\sim$250\degr. Second, two knots bright both in {\it B\arcmin} and {\it K} 
with very little diffuse H$\alpha$ emission at PA=270\degr\ and 300\degr; this is 
where Kohno et al. (1999) find the leading secondary maximum of HCN distribution, 
and also the location of the leading secondary maximum of the 6 cm radio emission
(Saikia et al. 1994). Third, the \ion{H}{ii} region to the north, PA=342\degr, 
that is brightest in {\it B\arcmin}, H$\alpha$, HCN and 6 cm. Finally, the region just behind 
this one along the ring, partly associated with the CO peak from Kohno et al., 
and that might be predicted to be the next in the sequence of star formation. 
A similar, albeit somewhat less well defined, trend is observed in the south-eastern 
side.

In conclusion, a qualitative look at the colours suggests that there may
be two coexisting modes of star formation, continuous and bursting, and that the
bursting mode may be sequential from the spearhead backwards along the entrained 
material as it is shocked in the circumnuclear ring.

Some of the ideas proposed here are only suggested by the wealth of information 
indicated by the data, and a detailed quantitative analysis of the stellar
populations will be necessary to elucidate the validity of some of them. We 
are gathering spectroscopic data at other wavelengths to address this future analysis.

\subsection{Linear analysis of the kinematics}

By performing a standard linear analysis of the kinematics 
(e.g. Binney \& Tremaine 1987) it is possible to obtain the location 
of the main resonances, co-rotation (CR) and the inner Linblad 
resonances (ILRs). We use results from numerical simulations that 
place CR close to the bar semimajor axis (Athanassoula 1992), 
to infer a bar pattern speed of 3 km~s$^{-1}$ arcsec$^{-1}$ (32 km~s$^{-1}$ kpc$^{-1}$)
\footnote{Computing the intersection between the bar pattern speed and the curve
$\omega-\kappa/2$ (were $\omega(r)$ is the angular velocity and $\kappa(r)$ 
the epicyclic frequency) directly from the data points in the velocity curve is hampered 
by the dependence of $\kappa(r)$ on the derivative of the velocity, which 
implies that small errors and structure in the velocity as a function of radius 
translate into large fluctuations in $\kappa(r)$. In the case of NGC~6951 this 
problem is further exacerbated by the scarcity of data points in the region 
outside the circumnuclear ring, between 10 arcsec and 30 arcsec.}.
This implies the existence of two ILRs, at 2 arcsec (180 pc, inner ILR) 
and at 12 arcsec (1100 pc, outer ILR)
\footnote{The shallow radial dependence of $\omega(r)$ at this distance from 
the nucleus, and the steeper dependence closer to the nucleus, implies that a 
10\% or 20\% uncertainty in the position of CR does not affect significantly 
the location of the ILRs.}. 
From OVRO CO observations of the central part of the galaxy, Kenney et al. (1992) 
obtain values of 180 pc for the iILR and 460 pc for the oILR; this latter value is
a lower limit derived from their upper limit to the bar pattern speed of 
66 km~s$^{-1}$ kpc$^{-1}$, and thus is compatible with our results.

\subsection{Structure of the circumnuclear region}

The linear analysis of the kinematics indicates the existence of a possible iILR at 2 arcsec. 
Do we see a correlation between this dynamical feature and any morphological or 
spectroscopic properties?

Morphologically, the H$\alpha$, colour, and HST optical images all show structure within 
the inner 2 arcsec; i.e., it is not a point like nucleus. Friedli et al. (1996) have 
looked without success for the existence of a secondary bar, that could be possibly
associated with the existence of the iILR. There is no indication of such an inner 
bar either in our images. In fact, the HST optical image shows a clear inward spiraling
structure down to 0.5 arcsec radius (46 pc).

Spectroscopically, several diagnostics are spatially resolved within the inner 2 arcsec
radius, such as the [\ion{N}{ii}]/H$\alpha$ ratio (Fig.~\ref{n2ha}), the FWHM of both 
H$\alpha$ and [\ion{N}{ii}] (Fig.~\ref{fwhm}), and the equivalent with of the 
CaT lines indicating the presence of an important population of red supergiants in the 
nucleus (Fig.~\ref{ewCaT}). The gaseous and stellar velocity curves also show features 
within the iILR: for the three PA, the two stellar components merge at the iILR 
and are no longer distinguishable; 
the gas velocity curves at PA=84{\degr} and 138{\degr} flatten within a 1.5
arcsec radius,  while at PA=48{\degr} the gas velocity shows a peculiar
structure within a 2 arcsec radius, that is not present in the stellar
velocity. 

This peculiar structure along the minor axis could be due to an off-centered slit, 
to a gaseous nuclear outflow, or to a structure related to the iILR. 
The first possibility is ruled out because this velocity structure is present only in the
ionized gas but not in the stars, that implies that it is a physically
existing feature, given that the two spectral regions were observed
simultaneously and through the same slit. Of the two other possibilities, our
data do not have sufficient spatial resolution to make a more definitive
conclusion. However, we can make a few comments. A nuclear outflow could be
produced by the combined effect of the winds and supernova explosions of the
recent past generation of star formation indicated by the red supergiant stars
we have detected. Saikia et al. (1994) give a total flux for the nuclear beam
component of NGC~6951 at 6 cm of 1.1 mJy; they argue that this non-thermal
flux is most likely due to supernovae. This flux implies a luminosity of
0.47$\times10^{20}$ watt Hz$^{-1}$, that can be converted to a supernova rate
using the models of Colina \& P\'erez-Olea (1992), to give 0.002 yr$^{-1}$. A
similar calculation with the 20 cm radio flux in the nuclear component given
by Vila et al. (1990), that implies a luminosity of 1.03$\times10^{20}$ watt
Hz$^{-1}$, yields the same value for the SN rate. This is a low SN rate when
compared with bright nuclear  starbursts such as NGC~7714, for which
Gonz\'alez Delgado et al. (1999a) find 0.07 yr$^{-1}$,  but is qualitatively
consistent with the faintness of the nucleus in H$\alpha$. If a nuclear 
outflow were the explanation for the kinematic features in the nucleus, we
would only be seeing the approaching, blueshifted, side of the outflow with a
speed of $\sim$100 km~s$^{-1}$; the typical expansion velocity of a superwind
bubble blown by a nuclear starburst in dwarf galaxies is of the order of 50
km~s$^{-1}$ (Marlowe et al. 1995); but is significantly larger, a few 100
km~s$^{-1}$, for nuclear superwinds in Starburst galaxies (Gonz\'alez Delgado
et al. 1998). The receding side would be presumably occulted by a combination
of nuclear obscuration and spatial and spectral resolution.

A third possibility suggested by the data is that the nuclear gas dynamics 
inside the iILR is partially decoupled from that of the circumnuclear region 
which, in turn, is partially decoupled from that at larger scales. 
This would be a scenario of nested disks within disks, where
a circumnuclear disk accumulates mass from the outer main galaxy disk via the torques
produced by the bar, and becomes sufficiently massive to decouple from the main disk
dynamics at the oILR; part of this infall proceeds further into the nucleus, where a
similar rotating structure (a torus or a nuclear disk such as those seen in HST images,
Ferrarese \& Ford 1999; Ford et al. 1998) decouples from the circumnuclear disk at the
iILR. We notice that the orientation of the ring in NGC 6951 is leading with respect to the inflow
of material from the bar, i.e. it is advanced from the perpendicular to the bar, and this
relative orientation is seen in numerical simulations only when the system is relatively
old (see, for example, figure 3 of Byrd et al. 1994). In this context, it would be possible
to interpret the change of the velocity curve in the three PA, within the
inner 2 arcsec radius, as produced by rotation in a disk with a minor axis
different from the 48{\degr} of the main disk, and different from the
30{\degr} of the circumnuclear disk (cf. section 4.3); it would have a value
somewhere between 84{\degr} and 48{\degr}, so that at PA=48{\degr} we could
observe the change of velocity slope with respect to that at 84{\degr}. 

The appearance of the HST images would support this scenario. Figure~\ref{HST2} shows 
the HST {\it V} and {\it H} sharp divided\footnote{This process of enhancement consists in dividing 
the original image by a median filtered version, so as to enhance sharp features.} images; 
the {\it H} image shows a very uniform stellar light distribution in the transition zone between 
the ring and the nucleus. However, the {\it V} image shows a structure spiralling
inwards, delineated by the dust that can be traced right into the nuclear 0.5
arcsec. Thus, these  high resolution images show both the lack of any inner
bar-like structure and the existing  of a continuing spiralling into the
nucleus, similar to that found in numerical simulations  by Piner et al.
(1995, cf. their fig.~4). A similar conclusion has been recently reached by 
Regan \& Mulchaey (1999), who have looked for evidences of strong nuclear bars
in WFPC2 and  NICMOS2 images of a sample of Seyfert galaxies, and find
evidence of the existence of this  nuclear bar in only 3 out of 12 galaxies
studied, while the majority of the galaxies show a  spiral morphology similar
to what we find in NGC~6951. This does not necessarily implies that the 'bars
within bars' mechanism proposed to fuel the nucleus has not been at work in 
these galaxies (Shlosman et al. 1989; Friedli \& Martinet 1993); at least in
NGC~6951 we  have found several indications that point to a system that is
dynamically old, in particular,  the large percentage of molecular mass
accumulated in the nucleus, $\sim25$\%, could  have already occasioned the
dissolution of a nuclear bar, and be working towards  the desassembling of the
large scale bar. 

This idea is inferred from extrapolations of three slit PA. While we understand that 
a complete 2D spectroscopic mapping of this and other galaxies is required to confirm
or otherwise dismiss it, it is also true that such a 2D mapping at a spectral
{\em and} spatial resolution equal to or better than our data is very difficult
to obtain with present day instrumentation. Recent developments like SAURON
(Miller et al. 1999) and similar  instruments will have an important impact on
such studies in the next few years.

\section{Summary and conclusions}

We have obtained broad band {\it B{\arcmin}IJK$\rm_s$} images and high resolution 
high dispersion longslit spectroscopy for the ionized gas (around H$\beta$ to 
[\ion{O}{iii}] and H$\alpha$ to [\ion{S}{ii}]), and for the stellar populations 
(in the \ion{Ca}{ii} triplet  lines around 8500 \AA) of the galaxy NGC~6951. 
Together with archival HST {\it VJH} images, we analyse these data to study the central
structure and kinematics and find that: 

\begin{itemize}
\item There is little star formation ongoing inside the bar dominated region of the
galaxy, except for the circumnuclear ring at 5 arcsec radius. There is some evidence that
this star formation occurs in two modes, in bursts and continuously, along the ring and 
inwards towards the nucleus. 
\item The nuclear spectrum shows both very high excitation and very strong [\ion{N}{ii}]
and [\ion{S}{ii}] lines, making the classification of the AGN somewhat uncertain between
a high excitation LINER and a possibly high nitrogen abundant Seyfert 2, depending mainly 
on the uncertain correction for underlying absorption in H$\beta$. The electron density 
varies between 300 cm$^{-3}$ and 1000 cm$^{-3}$. 
\item The equivalent width of the \ion{Ca}{ii} triplet absorption lines show that in the 
metal rich central region of this galaxy, within 5{\arcsec} radius, the
continuum light  is dominated by a population of red supergiant stars, while
outside the circumnuclear  ring the stellar population is that of giants. 
\item We suggest that the gaseous and stellar kinematics along the three slit position angles
can be interpreted as the existence of a hierachy of disks within disks, 
with dynamics decoupled at the two ILRs, that we find to be located at 180 pc and 
at 1100 pc. This would be supported by the structure seen in the high resolution HST images.
\item The nucleus is partly resolved within a radius of 1.5{\arcsec} (just inside the iILR)
both in the emission line ratio [\ion{N}{ii}]/H$\alpha$, and in the FWHM of the emission lines. 
\item Outside the iILR the stellar CaT velocity profile can be partly resolved into two different
components that seem to be associated to the bar and to a disk. 
\item We discuss the possibility that the kinematic component inside the iILR 
could be due to a nuclear disk, as in the disk within disk scenario suggested above, 
or to a nuclear outflow produced by the combined effects of SN and SN remnants.
\item Several clues indicate that this is a dynamically old system: (i) there is little star
formation ongoing inside the bar dominated part of the galaxy (except for the
circumnuclear ring), (ii) the relative amount of molecular to total mass within the inner 6 arcsec
radius is very large $\sim25$\%, and (iii) the geometry of the circumnuclear
ring leading at a position angle greater than 90{\degr} from the stellar bar.
It is thus possible that  a nuclear bar has existed in NGC~6951 that drove the
gas towards the nucleus, as in the  bars within bars scenario, but that this
bar has already dissolved by the gas accumulated  within the circumnuclear
region.

\end{itemize}

\begin{acknowledgements}

Thanks to Hector Aceves for precise comments and clarifications, 
and to Jaime Perea for SIPL.
This work is financed by DGICyT grants PB93-0139 and PR95-329. 
We acknowledge financial support from INSU-CNRS for several observing
trips and from the Picasso program of the French Ministry of Foreign
Affairs for several collaboration trips.
EP thanks the Director of the Space Telescope Science Institute for the occasion 
to visit the Institute during the course of this research. The Space Telescope 
Science Institute is operated by AURA, Inc., under NASA contract NAS5-26555. 
We are very grateful to Ron Probst, who made available to us the SQIID package 
for the reduction of infrared images within IRAF.
The BARS project has received observing time under the International
Time Programme offered by the CCI of the Canary Observatories, and
financial support by the European Commission through the Access to
Large-Scale Facilities Activity of the Human Capital and Mobility
Programme.  
This research has made use of the NASA/IPAC extragalactic database (NED), 
which is operated by the Jet Propulsion Laboratory, Caltech under contract 
with the National Aeronautics and Space Administration.

\end{acknowledgements}


\newpage 

\begin{figure}
\caption{
Full frame {\it B\arcmin} (a) and {\it Ks} (b) broad band images of NGC~6951. 
The images show clearly two spiral arms, the bar, dust lanes along the bar and the 
circumnuclear ring at 5 arcsec from the nucleus. The orientation in all the images is 
North up and East to the left. The scale is 92 pc arcsec$^{-1}$. The contour levels
are (a) in B{\arcmin} 24.0 22.5 21.5 20.5 19.5 19.0, (b) in {\it Ks} 21.0 20.0 19.5 19.0 
18.5 18.0 17.5 17.0 16.5 16 15.5 15.0.
}
\label{imBK}
\end{figure}

\begin{figure}
\caption{
Broad band photometric parameters as a function of the root square of the distance 
to the nucleus. Surface brigthness ($\mu$): upper pannels; ellipticity ($\epsilon$): middle
pannels; and position angle (PA): bottom pannels. The parameters are derived by fitting
isophotal profiles to the images: {\it B\arcmin} (left pannels); {\it I} (middle left pannels); 
{\it J} (middle right pannels) and {\it Ks} (right pannels).
}
\label{parBIJK}
\end{figure}

\begin{figure}
\caption{
Full frame colour images. (a) {\it B{\arcmin}-I} and (b) {\it J-Ks}. {\it B{\arcmin}-I} 
traces very well the spiral arms and the circumnuclear ring, where recent star formation 
is taking place. Dust lanes along the bar are also seen as dark features. Grayscales 
are displayed between {\it B{\arcmin}-I}=3.1 and 1.9, and {\it J-Ks}=-1.0 and -2.0.
} 
\label{NOTBINOTJK}
\end{figure}

\begin{figure}
\caption{
colour images of the bar and inner regions of NGC~6951: (a) {\it B\arcmin-K{\arcmin}} and 
(b) {\it I-K{\arcmin}}. The inner part of the ring is dominated by blue colours, except for 
the nucleus that appears as a red central knot.
Grayscales are displayed between {\it B{\arcmin}-K{\arcmin}}=0.06 and 0.16, 
and {\it I-K{\arcmin}}=0.55 and 1.30.
}
\label{BKIK}
\end{figure}

\begin{figure}
\caption{ {\it J-H} colour image of the central 12{\arcsec}$\times$12{\arcsec}. 
These images were observed by HST+NICMOS. The spatial scale is 0.075 {arcsec} pixel$^{-1}$,
and the grayscale has been displayed between {\it J-H}=0.55 and 0.31 mag.
The circumnuclear ring is mainly delinated as two short spiral arms. In the inner part of 
the ring, the dust delineates a multi-spiral structure that ends in the nucleus.}
\label{HST1}
\end{figure}

\begin{figure}
\caption{{\it B\arcmin-K\arcmin} colour image of the central 20$\times$20 arcsec. 
H$\alpha$ contours (from the data of Gonz\'alez Delgado et al. 1997) are also plotted. 
Note that there is not a perfect correspondence between the blue knots in 
the {\it B\arcmin-K{\arcmin}} image and the nebular emission. 
The three slit positions for the spectroscopy data are also plotted.
Grayscale is displayed between {\it B{\arcmin}-K{\arcmin}}=0.05 and 0.16.
}
\label{BKHanucleo}
\end{figure}

\begin{figure}
\caption{H$\alpha$/H$\beta$ emission line ratios (dots) measured along the three slit PA. 
This ratio has been corrected for underlying stellar absorption as explained in the text. 
The theoretical ratio predicted by the case B recombination is plotted as a horizontal 
dashed line. The H$\alpha$ flux along the slit is also shown as a full line.}
\label{hahb}
\end{figure}

\begin{figure}
\caption{H$\alpha$ (open circles + dashed line) and [\ion{N}{ii}] (squares + dashed) fluxes 
and their ratio [\ion{N}{ii}]/H$\alpha$ (filled circles + full line) measured along the 
three slit PA. Note that the large [\ion{N}{ii}]/H$\alpha$ ratio in the inner $\pm$2 arcsec 
is due to a strong increase in the [\ion{N}{ii}] flux.}
\label{n2ha}
\end{figure}

\begin{figure}
\caption{H$\beta$ (filled circles + dashed line) and [\ion{O}{iii}] (open points + full line)  
fluxes along PA=138\degr. The [\ion{O}{iii}] emission is mainly concentrated in the inner 
$\pm$2 arcsec, where the H$\beta$ emission is very weak.}
\label{hbo3}
\end{figure}
                                                   
\begin{figure}
\caption{One dimensional spectrum, from 4750 \AA\ to 5150 \AA, of the nucleus and of 
the two regions on either side of the nucleus along PA=48\degr. Note that these regions 
have low excitation, consistent with the gas being photoionized by stars.}
\label{hbspec}
\end{figure}

\begin{figure}
\caption{
Nuclear spectrum (the central 0.72 arcsec) in the \ion{Ca}{ii} triplet region. 
The most relevant lines are labelled. \ion{Mg}{i} $\lambda$8807 is very strong, indicating 
that the metallicity is solar. Paschen 14 is very weak, thus Pa13 and 
Pa15 must contribute very little to the \ion{Ca}{ii} $\lambda$8542 and $\lambda$8662, 
respectively.
}
\label{CaTspec}
\end{figure}

\begin{figure}
\caption{
Equivalent width of the \ion{Ca}{ii} $\lambda$8542 + $\lambda$8662 lines,
 measured using the line and continuum windows defined by D\'\i az et al. (1989). 
The equivalent width is $\geq$ 7 \AA\ in the nucleus and out to the circumnuclear ring,
indicating the presence of red supergiant stars in the central 5$\times$5 arcsec of the galaxy.
}
\label{ewCaT}
\end{figure}

\begin{figure}
\caption{
Velocity curve (dots) obtained from the pixel to pixel measurements of the H$\alpha$ 
flux (full line), derived fitting a gaussian to the profile of the emission line. Curves are 
plotted for the three slit positions. At PA=138{\degr} (photometric major axis of the galaxy) 
the curve shows the maximum velocity amplitud. At PA=48{\degr} the curve shows more 
structure than expected for the minor axis of the galaxy. The inset within each figure shows the 
circumnuclear $\pm8$ arcsec expanded; the three insets have the same scales in both distance and
velocity axes, respectively.
}
\label{veloflux12}
\end{figure}
 
\begin{figure}
\caption{FWHM of the emission lines H$\alpha$ (filled circles) and [\ion{N}{ii}] 
(empty squares). The velocity dispersion of the gas is maximum in the inner $\pm$1.2 arcsec, 
and minimum at the position of the \ion{H}{ii} regions in the ring. The H$\alpha$ flux is 
plotted as a dashed line.}
\label{fwhm}
\end{figure}

\begin{figure}
\caption{
Velocity curves from the two strongest CaT absorption lines 
($\lambda\lambda$8542, 8662), measured by cross-correlation of the galaxy frames with spectra 
of K giant stars. The velocity curve of the ionized gas (dots) is shown for comparison with
the stellar velocity curve (circles). In the central region the CaT absorption lines are 
resolved in two components, plotted as filled and open circles. The insets show the central
$\pm$10 arcsec; the three have the same radial and velocity scales respectively, and the
gas velocity has been plotted as a full line. Notice how in the inner $\pm$2 arcsec, 
the stars and the ionized gas follow the same velocity pattern at PA=84{\degr} and 138{\degr}, 
but not along the minor axis, where the H$\alpha$ curve presents more structure.
}
\label{veloCaT12}
\end{figure}

\begin{figure}
\caption{
HST WFPC2 {\it V} (a) and NICMOS {\it H} (b) images of the circumnuclear region. 
To show more clearly the morphological structure of the cicumnuclear region, the original 
images have been sharp enhanced by dividing them by the corresponding median filtered image. 
Note the spiralling structure delineated by dust in the {\it V} image down into the nucleus.
}
\label{HST2}
\end{figure}

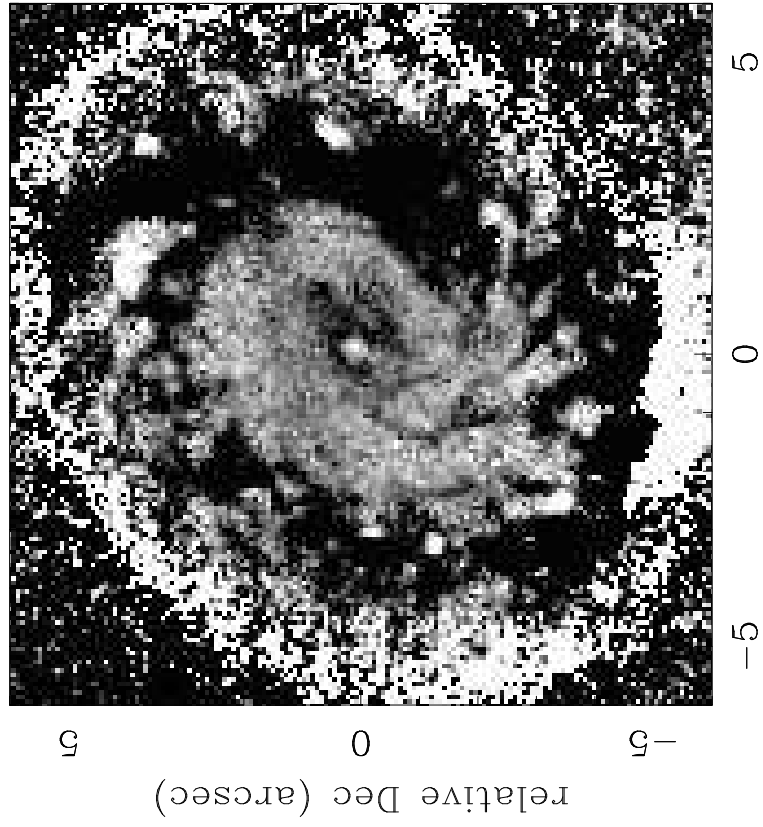
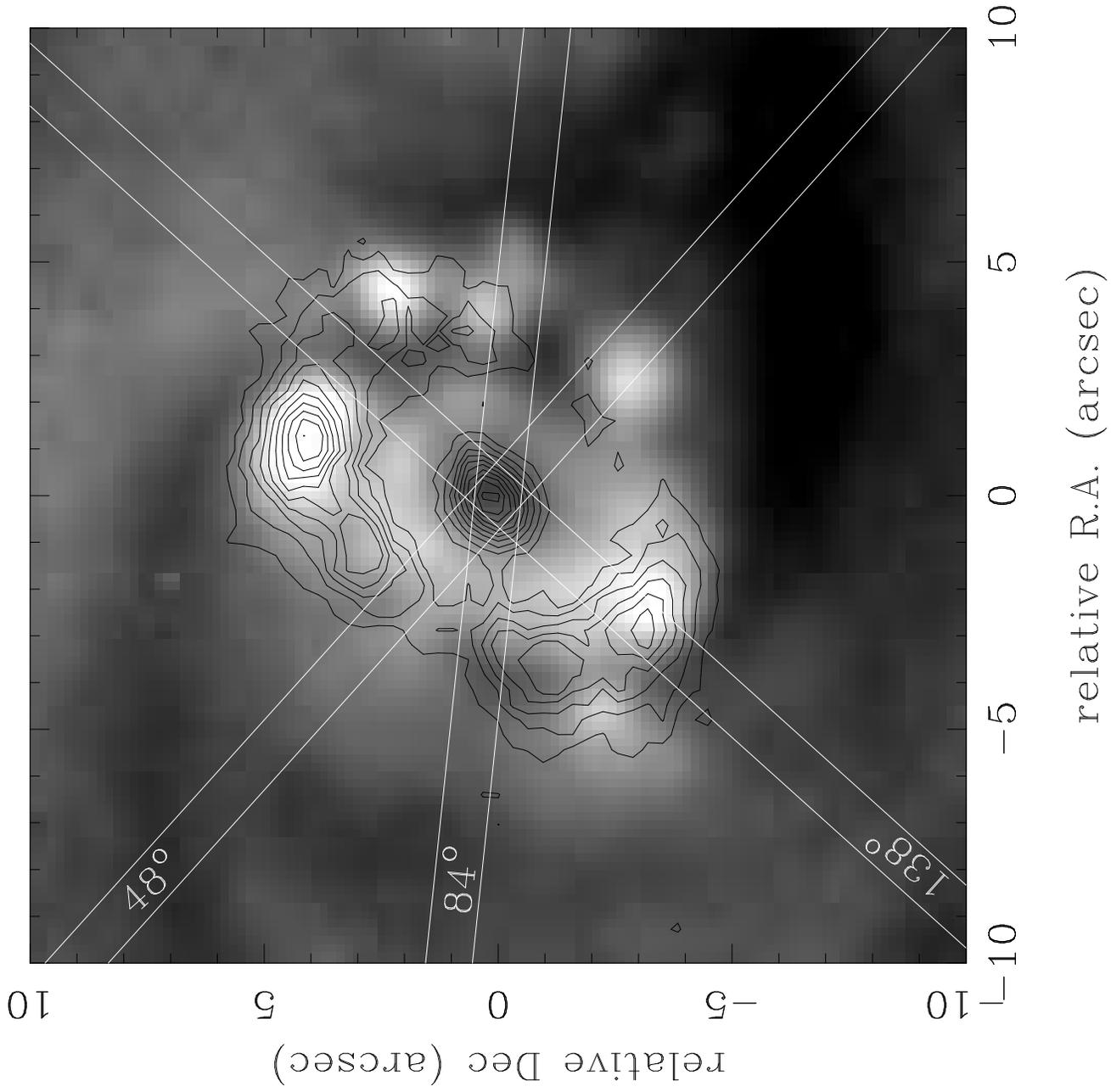
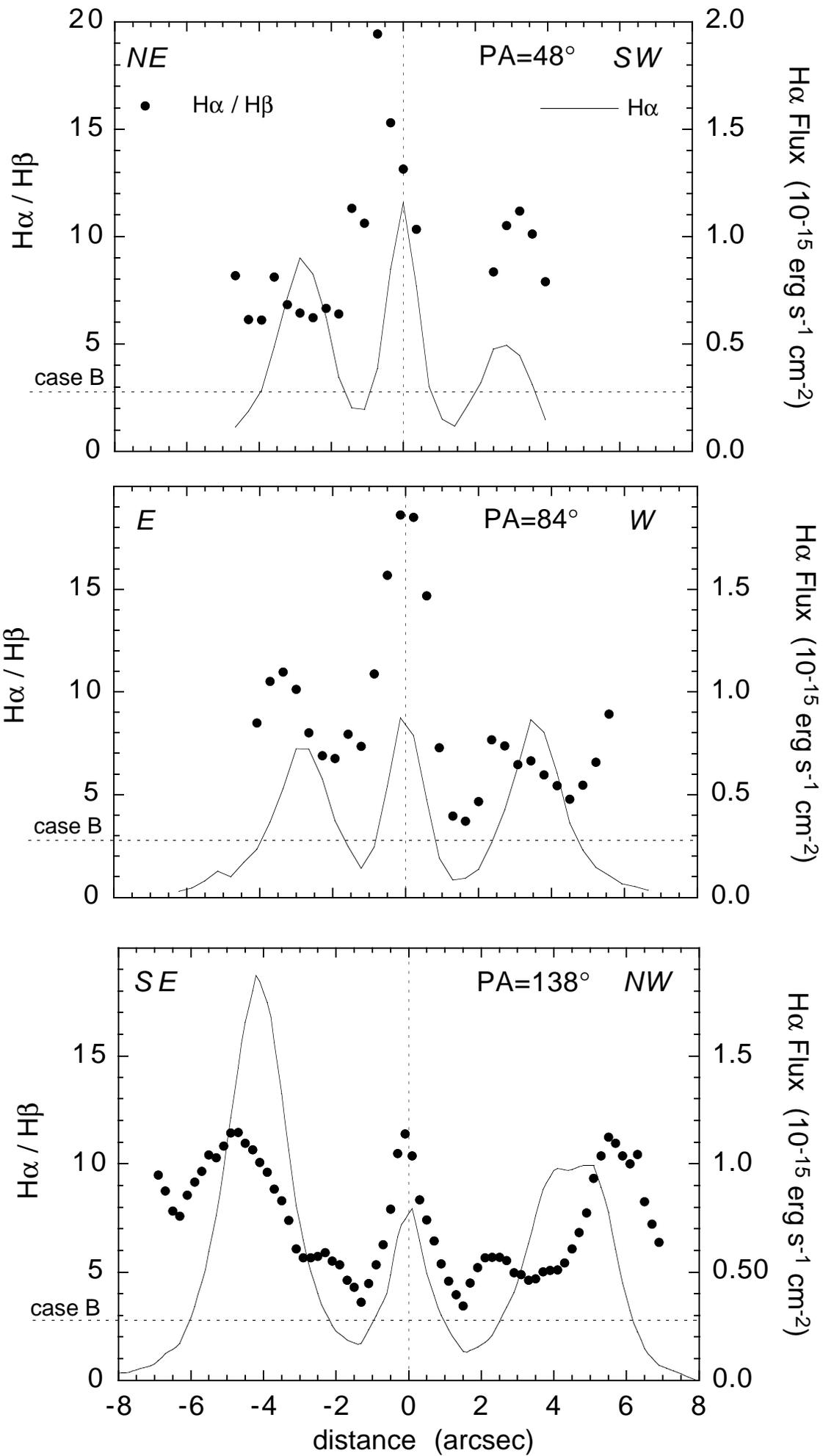
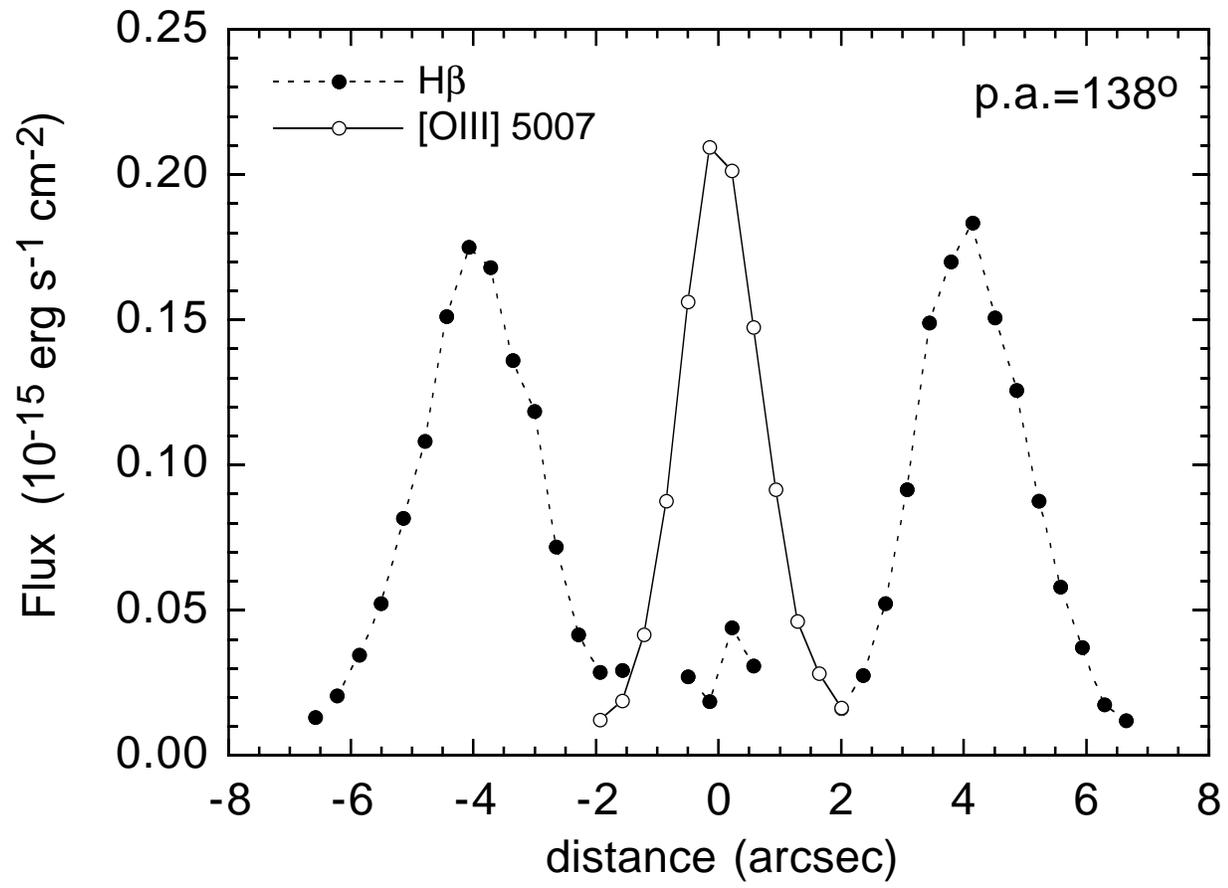
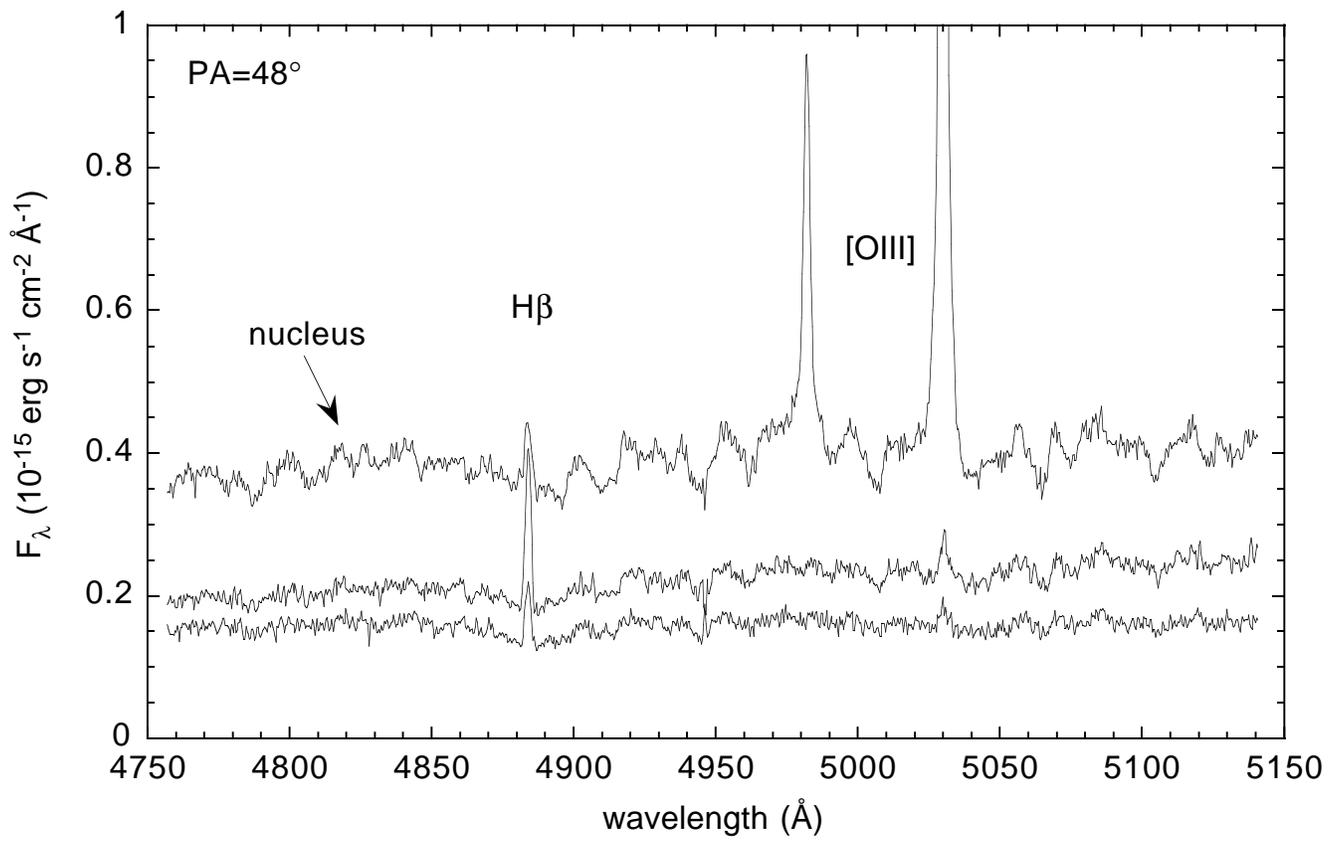
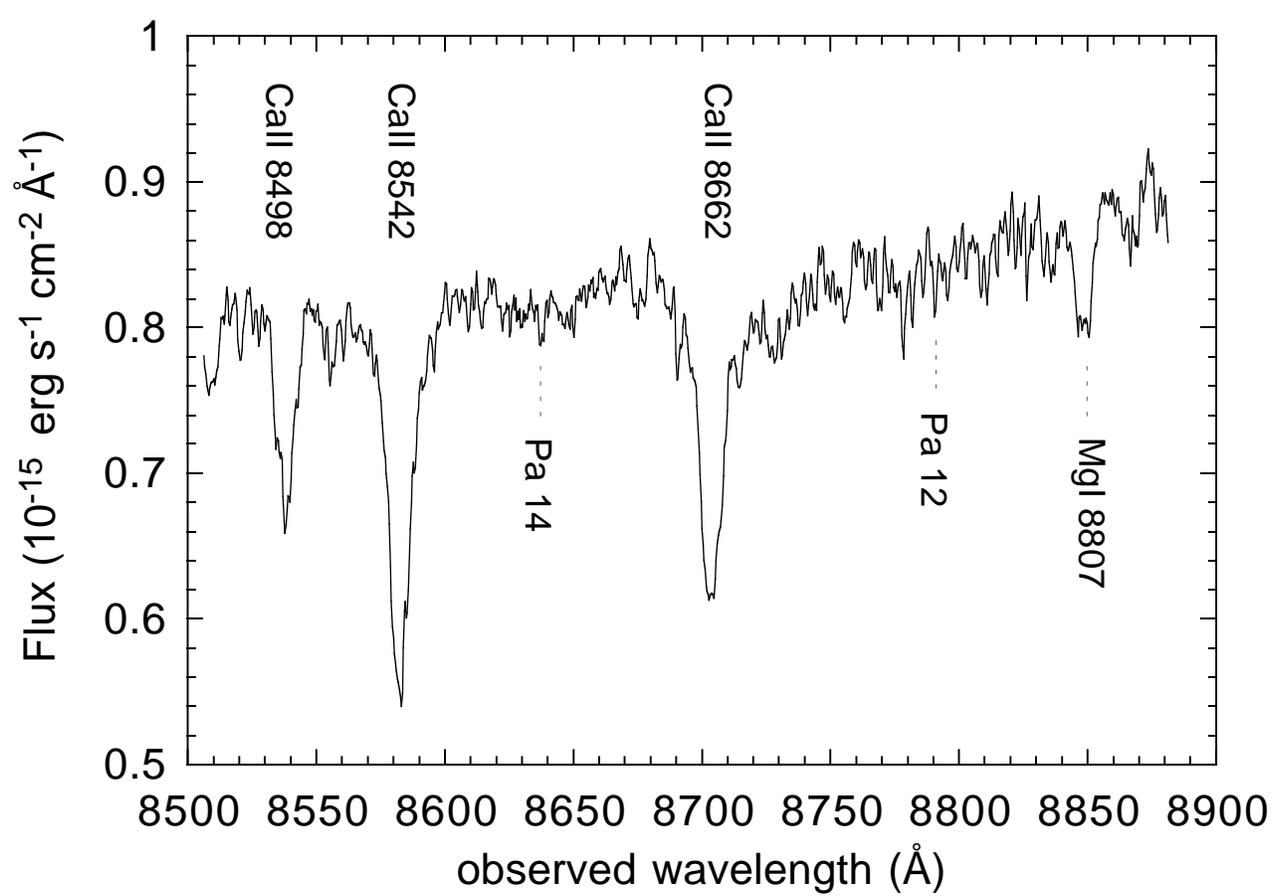
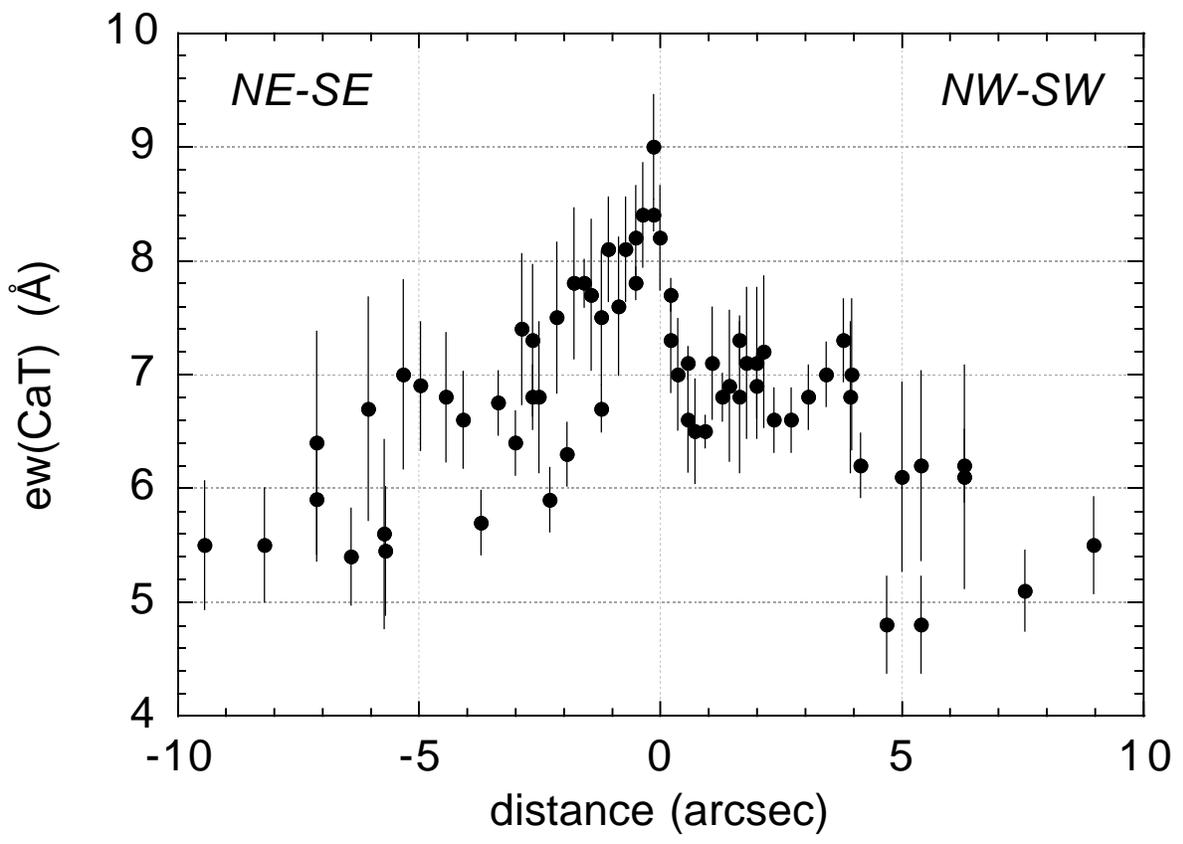
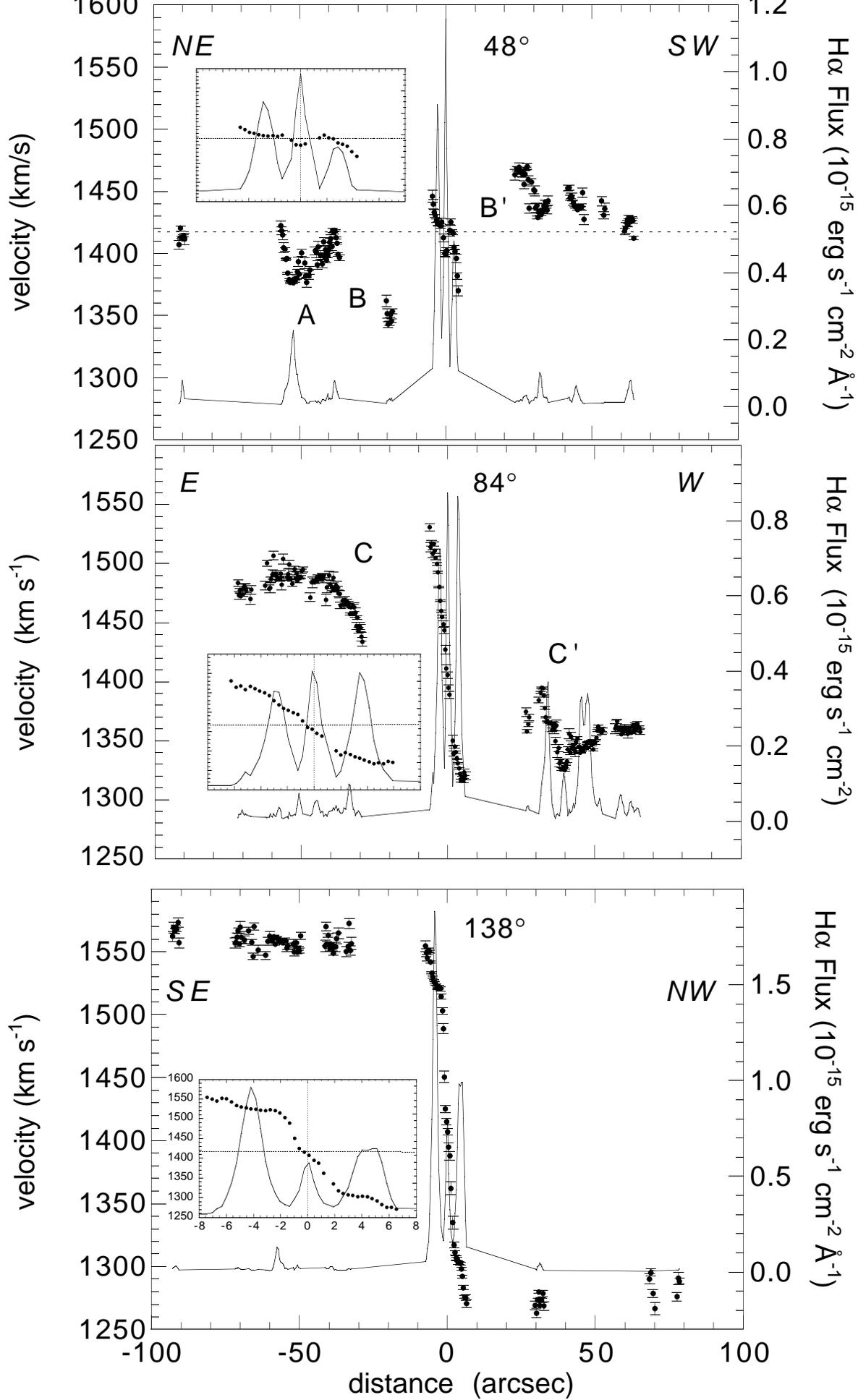
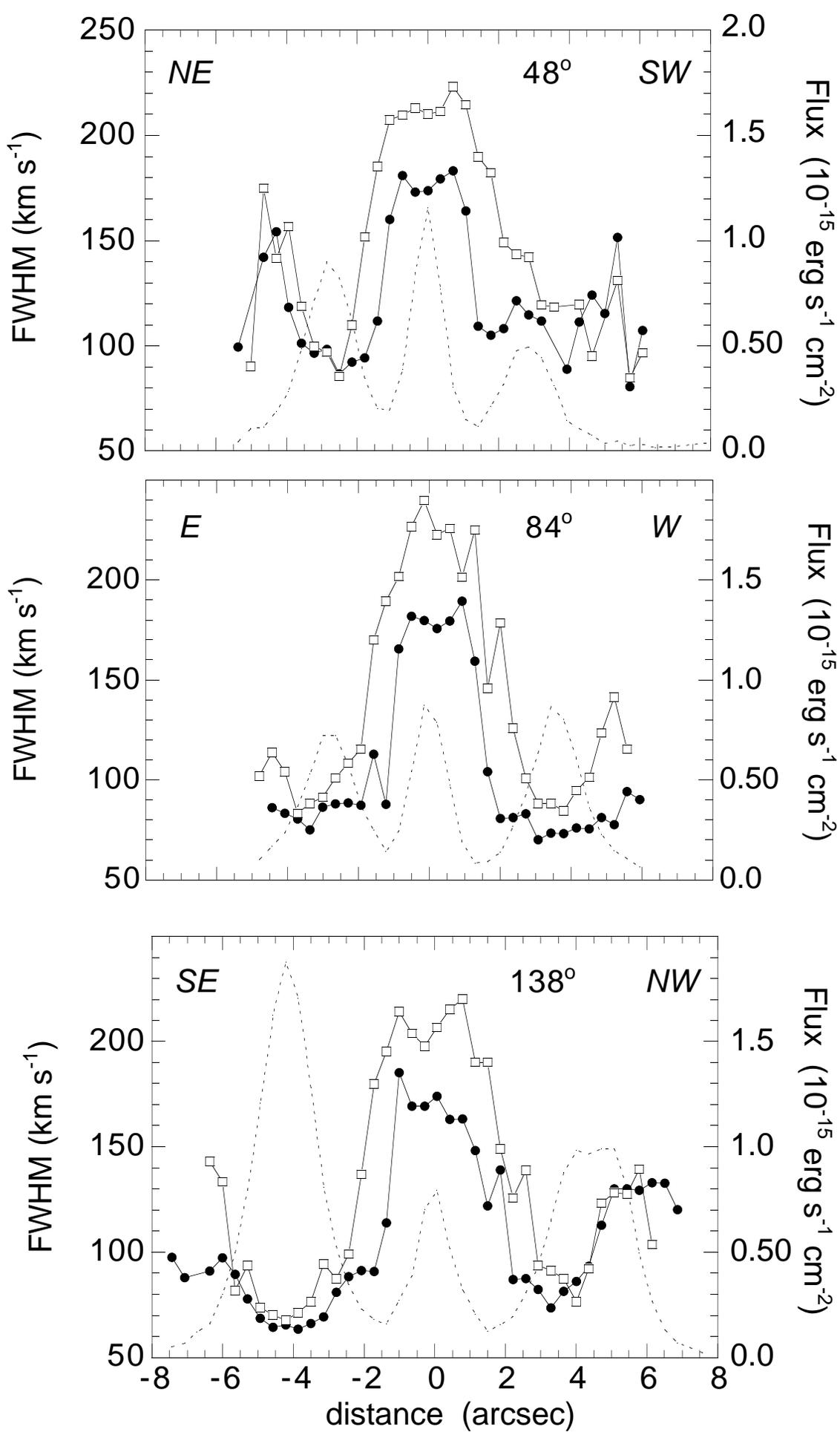
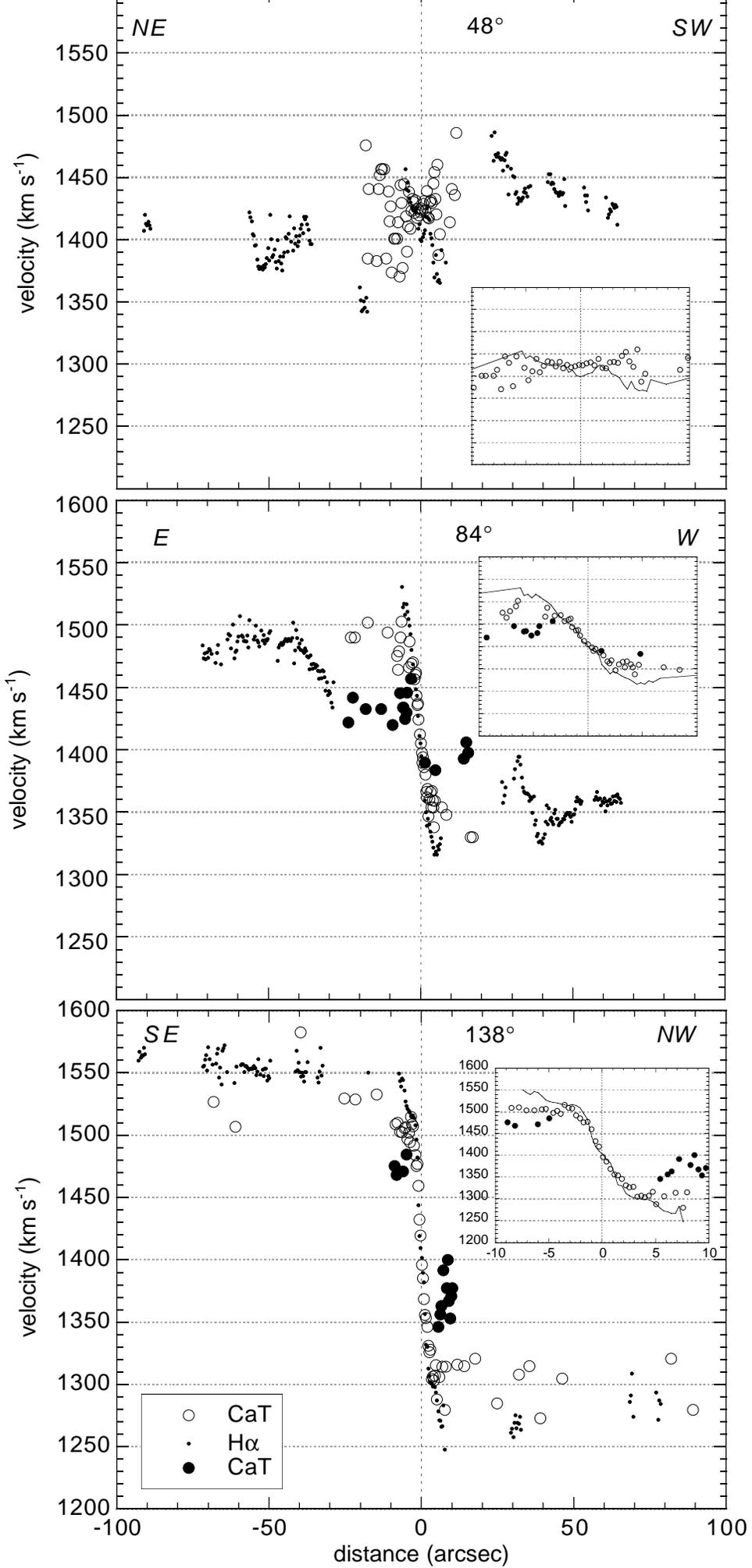
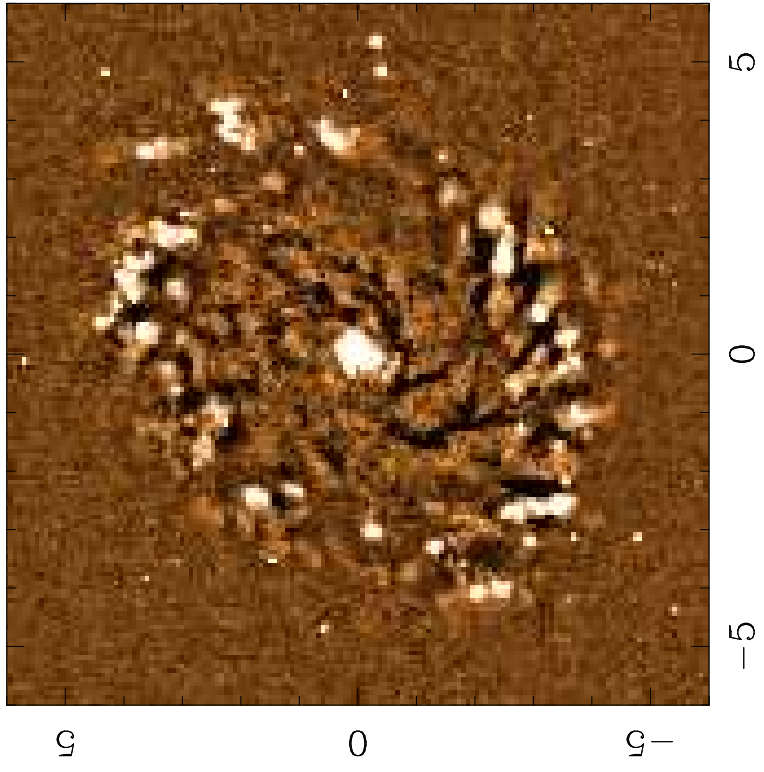
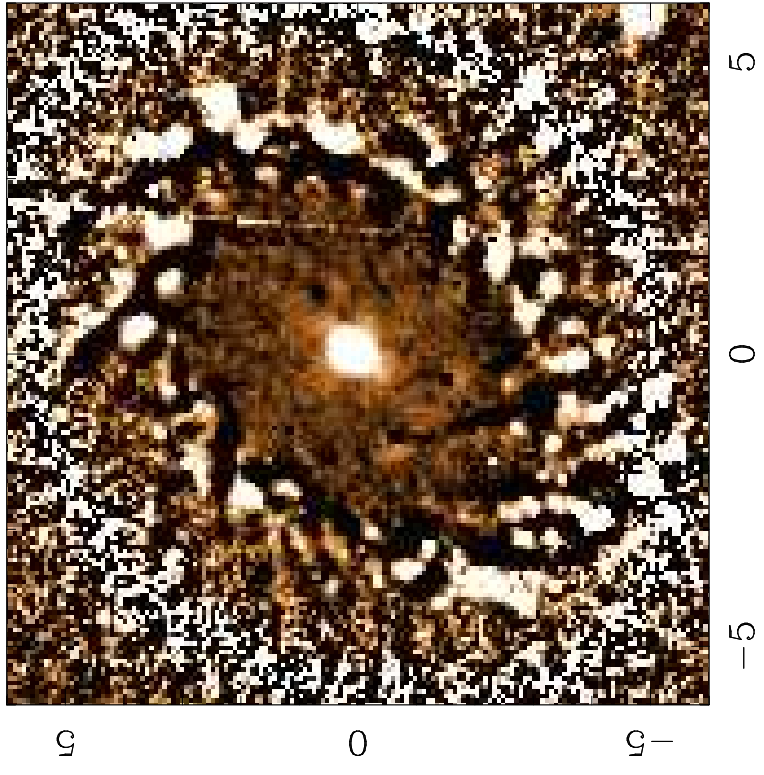
\end{document}